\documentclass[%
 reprint,
 amsmath,amssymb,
 aps,
]{revtex4-2}

\usepackage{graphicx}
\usepackage{dcolumn}
\usepackage{bm}
\usepackage{physics}
\usepackage{mathtools}

\begin{document}

\preprint{APS/123-QED}

\title{Impact of the nuclear charge distribution on the $g$-factors and ground state energies of bound muons}

\author{Atakan Çavuşoğlu}
\affiliation{Max Planck Institute for Nuclear Physics, Saupfercheckweg 1, 69117 Heidelberg, Germany}
\affiliation{Department of Physics, Boğaziçi University, Istanbul, Turkey}
\affiliation{Department of Electrical and Electronics Engineering, Boğaziçi University, Istanbul, Turkey}

\author{Bastian Sikora}
\affiliation{Max Planck Institute for Nuclear Physics, Saupfercheckweg 1, 69117 Heidelberg, Germany}

\date{November 28, 2023}

\begin{abstract}
    The finite nuclear size corrections to the ground state energies and $g$-factors in muonic atoms are investigated for several elements. Both approximative and exact solutions of the one-particle Dirac equation with both the homogeneous sphere nucleus model and the Fermi distribution nucleus model are presented, and the leading nuclear deformation effects on the $g$-factors and ground state energies are also evaluated. The electronic, muonic, and hadronic electric-loop vacuum polarization corrections are calculated for point-like, spherical, and Fermi nucleus models. The obtained results show a heavy dependence on the chosen nuclear model, and highlight the importance of constructing precise theoretical models for the nucleus for accurate QED predictions of the observables of muonic atoms.
\end{abstract}

\maketitle

\section{Introduction}
In comparison to the electron, the muon's $\sim$207 times heavier mass renders bound muonic systems an ideal testing field for investigating the nuclear effects on bound particles. Qualitatively, the orbital radii of bound particles are inversely proportional to their mass~\cite{Greiner1997}, which, in the case of the bound muon, causes the muonic orbit radii to be comparable to the nuclear radii, which significantly amplifies the nuclear effects on the properties of the bound muon in comparison to the electron, especially for muons bound to heavier nuclei.

Nuclear effects on the $g$-factors of bound electrons have been a major theoretical \cite{Beier2000,Shabaev2006,Yerokhin2013,Yerokhin2016,Sikora2020,Cakir2020,Debierre2020,Valuev2020,Debierre2021} and experimental \cite{Sturm11,Sturm13,Sturm14,Arapoglou2018,Sailer2022,Schneider2022,Morgner2023} research interest in the past years, and significant progress has been made in determining the nuclear contributions to the energy spectra of muonic atoms \cite{Wu1969,Michel2017,Patoary2018,Michel2019_2,Paul2021,Okumura2021,Valuev2022,Oreshkina2022}, which has been a major tool in determining the properties of various nuclei \cite{Schaller1980,Ruetschi1984,Piller1990,Pohl2010,Antognini2021,Saito2022}. Nevertheless a thorough investigation of the nuclear corrections to the bound muon $g$-factor still remains untouched, with the exception of the very light muonic $^4$He$^+$ ion~\cite{Sikora2018}.

In the present work, we investigate both the direct nuclear effects on the ground state energies and $g$-factors in muonic atoms, as well as the nuclear corrections to the one-loop vacuum polarization contributions. We consider heomogeneous sphere, two-parameter Fermi, and deformed Fermi charge distribution models for the nucleus, and we calculate the first-order electronic, muonic, and hadronic electric-loop vacuum-polarization corrections. Throughout this work we use muonic natural units with $m_{\mu}=c=\hbar=1$ and the charge units $\alpha = e^2/4\pi$.

\section{Dirac Equation for Central Potentials}

The stationary states of a bound fermion can be obtained from the solutions of the time independent Dirac equation, which is the relativistic wave equation for spin-1/2 particles:
\begin{equation} \label{diracEq}
    \left[ \vb*{\alpha} \vdot \vb{p} + M \beta + V(\vb{r}) \right] \psi(\vb{r}) = E \psi(\vb{r}),
\end{equation}
where $\vb*{\alpha}$ and $\beta$ are Dirac matrices and $M$ is the mass of the bound particle. The eigenvalues $E$ of this equation are the bound state energies, and the eigenfunctions $\psi(\vb{r})$ are bound state wavefunctions.

In the case of a spherically symmetric potential, i.e. $V(\vb{r}) = V(r)$, the wavefunctions can be separated into radial and angular components \cite{Greiner1997}
\begin{equation}
    \psi(\vb{r}) = \begin{pmatrix}
        \frac{i}{r} G(r) \Omega_{\kappa m_j}(\theta, \phi) \\
        -\frac{1}{r} F(r) \Omega_{-\kappa m_j}(\theta, \phi)
    \end{pmatrix},
\end{equation}
where $\kappa$ is the relativistic angular momentum quantum number with $\kappa = -l-1$ for $j=l+1/2$ and $\kappa = l$ for $j=l-1/2$, and $m_j$ is the z-component of the total angular momentum $\vb{J}=\vb{L}+\vb{S}$. In the case of the $1s$ (ground) state, $\kappa = -1$ and $m_j = 1/2$. $\Omega_{\kappa m_j}$ are the spherical spinors. $G(r)$ and $F(r)$ are the solutions of the radial part of the Dirac equation (\ref{diracEq}):
\begin{equation} \label{radialDirac}
\begin{split}
    \dv{G(r)}{r} + \frac{\kappa}{r} G(r) - (M - V(r)) F(r) &= E F(r), \\
    -\dv{F(r)}{r} + \frac{\kappa}{r} F(r) + (M + V(r)) G(r) &= E G(r).
\end{split}
\end{equation}

In the case of the point-like nucleus, the potential is the Coulomb potential 
\begin{equation} \label{coulomb}
    V_{\text{C}}(r) = -\frac{Z \alpha}{r},
\end{equation}
and the eigenfunctions and eigenvalues of the radial Dirac equation (\ref{radialDirac}) are known analytically (see \cite{Greiner1997}). The energy eigenvalues express the fine structure of the energy spectra of atoms:
\begin{multline} \label{E0}
    E_{\text{C}} = \\ M \left[ 1 + \frac{(Z\alpha)^2}{\left[n - j - \frac12 + \left[\left(j + \frac12\right)^2 - (Z\alpha)^2 \right]^{1/2}\right]^2} \right]^{-1/2}
\end{multline}

\subsection{$g$-Factor of a Bound Dirac Particle}
The so-called $g$-factor characterizes the coupling strength of a particle to an external magnetic field. For a bound Dirac particle in a state $\ket{\psi(\vb{r})}$, to first order in perturbation theory, it is defined by the relation
\begin{equation}
    \frac{e \vb{B}\vdot \expval{\vb{J}}}{2M} g = e \bra{\psi(\vb{r})} \vb*{\alpha} \vdot \vb{A} \ket{\psi(\vb{r})},
\end{equation}
where $\vb{A}$ is the vector potential corresponding to the constant magnetic field $\vb{B}$, i.e. $\vb{A} = 1/2 (\vb{B} \cross \vb{r})$. For a spherically symmetric potential and a particle in a stationary state, using the properties of the spherical spinors, it can be shown that (see \cite{Rose1961})
\begin{equation} \label{gFactorIntegral}
    g = \frac{2 M \kappa}{j (j+1)} \int \dd{r} r G(r) F(r),
\end{equation}
which, by using the radial Dirac equations (\ref{radialDirac}) and the mass derivative of the Hamiltonian $\pdv*{H}{M} = \beta$, can be expressed as \cite{Karshenboim2005}
\begin{equation} \label{gFromMassDer}
    g = -\frac{\kappa}{2j(j+1)} \left[1-2\kappa \pdv{E}{M}\right], 
\end{equation}
for any central potential that is independent of the mass of the particle. For a particle in the ground state, $\kappa = -1$ and $j = 1/2$, so the formula simplifies to
\begin{equation}
    g = \frac23 \left( 1 + 2 \pdv{E}{M}\right).
\end{equation}

If we express any central potential as a deviation from the Coulomb potential
\begin{equation}
    V(r) = V_C(r) + \Delta V(r),
\end{equation}
the ground state energy for this potential can also be expressed as a deviation from the Coulomb ground state energy
\begin{equation}
    E = E_C + \Delta E,
\end{equation}
with $E_C = M \sqrt{1 - (Z\alpha)^2}$. As a result, the deviations of the $g$-factors from the Coulomb $g$-factor can also simply be expressed as
\begin{equation}
    g_C + \Delta g = \frac23 \left(1 + 2\sqrt{1 -(Z\alpha)^2}\right) + \frac43 \pdv{\Delta E}{M},
\end{equation}
and 
\begin{equation} \label{deltaGFromMassDer}
    \Delta g = \frac43 \pdv{\Delta E}{M}.
\end{equation}

\section{Finite Nuclear Size Corrections}
One of the simplest nuclear models that take the finite size of the nucleus is the homogeneously charged sphere, with the charge distribution
\begin{equation} \label{sphereDistribution}
    \rho_{\text{Sphere}}(r) = \begin{cases}
        \frac{3Ze}{4\pi r_0^3} & \text{if } r \leq r_0 \\
        0 & \text{if } r > r_0
    \end{cases}
\end{equation}
where $r_0$ is the radius of the nucleus, related to the root-mean-squared radius of the nucleus by~\cite{Beier2000}
\begin{equation}
    r_0 = \sqrt{\frac{5}{3}} r_{\text{rms}}.
\end{equation}

For the homogeneously charged sphere, the potential outside the nucleus is the same as the Coulomb potential (\ref{coulomb}), but the potential inside the nucleus is different:~\cite{Patoary2018}
\begin{equation} \label{potentialSphere}
    V(r) = \begin{cases}
        -\frac{Z \alpha}{2r_0}\left(3-\frac{r^2}{r_0^2}\right) & \text{if } r \leq r_0 \\
        -\frac{Z \alpha}{r} & \text{if } r > r_0
    \end{cases}
\end{equation}

The radial Dirac equation (\ref{radialDirac}) can be solved for the region inside the nucleus ($r \leq r_0$) in power series, and since the potential (\ref{potentialSphere}) is a polynomial in $r$, a simple recursion relation can be obtained for the coefficients of the power series. The wavefunction inside the nucleus is \cite{Patoary2018}
\begin{equation} \label{wavefunctionSphereIn}
    \begin{pmatrix}
        G_{r<r_0}(r) \\ F_{r<r_0}(r)
    \end{pmatrix} = N_1 r^{\abs{\kappa}} \sum_{i=0}^{\infty} \left[a_i \pm (-1)^{i+1}\frac{\kappa}{\abs{\kappa}}a_i\right] r^i,
\end{equation}
with the recursion relation
\begin{equation}
    a_i = \frac{a_{i-1} \left[E + \frac{3Z\alpha}{2r_0} -M(-1)^i \frac{\kappa}{\abs{\kappa}} \right] -\frac{Z\alpha}{2r_0^3}a_{i-3}}{\kappa + (-1)^{i+1}\frac{\kappa}{\abs{\kappa}} \left(i + \abs{\kappa}\right)}.
\end{equation}

Since the potential outside the nucleus is the same as the Coulomb potential, the solution of the Dirac equation in this region follows similarly to the solution for the point nucleus (see \cite{Greiner1997}), with the difference being that the solutions singular at $r=0$ are also allowed. For a given E, the radial Dirac equation (\ref{radialDirac}) has two linearly independent solutions, but because of the normalizability of the wavefunctions, only the specific linear combinations that are regular at infinity are allowed. This condition reduces the number of possible solutions to one, and the wavefunction outside the nucleus can be written as \cite{Patoary2018}
\begin{multline} \label{wavefunctionSphereOut}
    \begin{pmatrix}
        G_{r>r_0}(r) \\ F_{r>r_0}(r)
    \end{pmatrix} = \frac{N_2}{\kappa + \frac{M Z \alpha}{\lambda}} \left( 2\lambda r\right)^{-\frac12} \sqrt{M \pm E} \\
    \times \left[ \left( \kappa + \frac{M Z \alpha}{\lambda} \right)W_{q,\gamma}(2\lambda r) \pm W_{q+1,\gamma}(2\lambda r) \right],
\end{multline}
where $\lambda = \sqrt{M^2 - E^2}$, $q = Z \alpha E/\lambda - 1/2$, $\gamma = \sqrt{\kappa^2 - (Z\alpha)^2}$, and $W_{a,b}(x)$ are the Whittaker functions of the second kind \cite{Abramowitz1965}.

The quantization of the energy is obtained from the continuity of the wavefunction at the edge of the nucleus,
\begin{equation}
    \begin{pmatrix}
        G_{r<r_0}(r_0) \\ F_{r<r_0}(r_0)
    \end{pmatrix} = \begin{pmatrix}
        G_{r>r_0}(r_0) \\ F_{r>r_0}(r_0),
    \end{pmatrix}
\end{equation}
or, alternatively,
\begin{equation} \label{sphereQuantizationCondition}
    \frac{G_{r<r_0}(r_0)}{F_{r<r_0}(r_0)} = 
    \frac{G_{r>r_0}(r_0)}{F_{r>r_0}(r_0)}
\end{equation}
which can be solved numerically to obtain the energy eigenvalues. After the eigenvalues are found, the normalization constants $N_1$ and $N_2$ in equations (\ref{wavefunctionSphereIn}) and (\ref{wavefunctionSphereOut}) can also be calculated numerically from the normalization condition of the wavefunction and the continuity condition at $r=r_0$.

After the ground state energy is found, it is also straightforward to find the finite size correction to the $g$-factor. By taking the mass derivative of the quantization condition in equation (\ref{sphereQuantizationCondition}) and using equation (\ref{deltaGFromMassDer}), we find
\begin{equation} \label{EnergyMassDer}
    \pdv{E_{Sph}}{M} =  - \frac{\pdv{M}(\frac{G_{r<r_0}(r_0)}{F_{r<r_0}(r_0)}-\frac{G_{r>r_0}(r_0)}{F_{r>r_0}(r_0)})}{\pdv{E}(\frac{G_{r<r_0}(r_0)}{F_{r<r_0}(r_0)}-\frac{G_{r>r_0}(r_0)}{F_{r>r_0}(r_0)})}
\end{equation}
and
\begin{equation}
    \Delta g_{\text{Sph}} = - \frac43 \frac{\pdv{M}(\frac{G_{r<r_0}(r_0)}{F_{r<r_0}(r_0)}-\frac{G_{r>r_0}(r_0)}{F_{r>r_0}(r_0)})}{\pdv{E}(\frac{G_{r<r_0}(r_0)}{F_{r<r_0}(r_0)}-\frac{G_{r>r_0}(r_0)}{F_{r>r_0}(r_0)})} - \frac43\sqrt{1-(Z\alpha)}.
\end{equation}
Having analytical expressions for the wavefunctions in the case of the homogeneous sphere model allows us to take the mass and energy derivatives analytically, and the corrections to the $g$-factors can be calculated exactly, without having to normalize the wavefunctions or take integrals of the form (\ref{gFactorIntegral}).

\subsection{Deviations from the Sphere Model}
Although the sphere model for the nucleus allows simple calculations of the finite nuclear size effects on the energy levels and $g$-factors, it is important to investigate the dependence of finite size corrections on the specific nuclear model used, especially for muonic atoms, such as more realistic charge distributions like the Fermi distribution or deformed Fermi distribution. Furthermore, quantum electrodynamic (QED) effects that can be expressed as an effective potential, such as one-loop vacuum polarization corrections, need to be investigated for extended nuclei rather than point-nuclei. As we will show in this paper, in the case of muonic atoms, the corrections due to these effects can be larger than the uncertainties to the finite size corrections due to the rms radius uncertainties, and therefore require a detailed consideration and understanding.

A method for estimating corrections for a wide range of nuclear charge distributions, by finding an effective radius and using the sphere model with that radius, was proposed in \cite{Shabaev1993}, and this method was applied to electronic atoms for calculating the finite nuclear size effects and nuclear deformation effects \cite{Karshenboim2005, Kozhedub2008, Zatorski2012, Michel2019}. But since the formulas derived in \cite{Shabaev1993} are based on the approximation that the nuclear radius is much smaller than the Bohr radius of the bound particle, they are not applicable to muonic atoms, and other approximation methods need to be used.

Here, we present a discussion of different methods for handling effects that can be represented as a correction to the sphere model potential, in which case the potential becomes
\begin{equation} \label{pertPotential}
    V(r) = V_{\text{Sph}}(r) + \delta V(r)
\end{equation}

In the following discussion, uppercase $\Delta$ will be used to denote the leading finite size correction to the point-nucleus case, and lowercase $\delta$ will be used to denote the corrections to the extended nucleus quantities.

\subsubsection{First Order Perturbation Theory}
If $\delta V(r)$ is small in comparison to $V_{\text{Sph}}(r)$, it can be regarded as a perturbation and first order perturbation theory can be applied, as
\begin{equation}
\begin{split}
    \delta E_{\text{pert}} &= \bra{\psi_{\text{Sph}}(\vb{r})} \delta V(r) \ket{\psi_{\text{Sph}}(\vb{r})} \\
    &= \int \dd{r} \left( G_{\text{Sph}}(r)^2 + F_{\text{Sph}}(r)^2 \right) \delta V(r).
\end{split}
\end{equation}
The perturbation integral can be evaluated numerically.

For calculating the corrections to the $g$-factor, equation (\ref{deltaGFromMassDer}) can be used. If $\delta V(r)$ is independent of the mass of the bound particle, the derivative with respect to mass only affects the wavefunctions:
\begin{multline} \label{perturbativeGFactor}
    \pdv{\delta E_{\text{pert}}}{M} = 2\frac{\bra{\psi_{\text{Sph}}} \delta V(r) \pdv{M}\ket{\psi_{\text{Sph}}}}{\bra{\psi_{\text{Sph}}}\ket{\psi_{\text{Sph}}}} \\
    - 2 \frac{\bra{\psi_{\text{Sph}}} \delta V(r)\ket{\psi_{\text{Sph}}} \bra{\psi_{\text{Sph}}}\pdv{M}\ket{\psi_{\text{Sph}}}}{\left[ \bra{\psi_{\text{Sph}}}\ket{\psi_{\text{Sph}}} \right]^2}
\end{multline}
This expression is valid for normalized, as well as for unnormalized wavefunctions, which avoids the calculation of the derivative of the normalization constant. In the derivatives of the wavefunctions, the derivatives of $E$ needs to be taken into account as well, which can be calculated from equation (\ref{EnergyMassDer}). Afterwards, the correction to the $g$-factor is simply
\begin{equation}
    \delta g_{\text{pert}} = \frac43 \pdv{\delta E_{\text{pert}}}{M}.
\end{equation}

\subsubsection{Variational Approximation}
Another approximation method is the variational method that is based on the fact that the expectation value of the Hamiltonian in any state other than the ground state is always greater than the ground state energy \cite{Sakurai}. This fact is exploited by choosing a trial wavefunction that is dependent on one or more parameters, and minimizing the expectation value of the Hamiltonian in that state with respect to the free parameters. In our case, we can choose the trial wavefunction as the ground state of the sphere nucleus Hamiltonian, with the radius of the nucleus left as a free parameter with respect to which the expectation value of the true Hamiltonian is to be minimized.

Specifically, the Hamiltonian for the sphere nucleus potential, with variable radius $R$, is
\begin{equation}
    H_{\text{Sph},R} = \vb*{\alpha} \vdot \vb{p} + M \beta + V_{\text{Sph},R}(r),
\end{equation}
and the ground state energy depends on $R$. The exact Hamiltonian of the system can be written as
\begin{equation}
    H = H_{\text{Sph},R} + H'_R,
\end{equation}
with
\begin{equation} \label{varPotential}
    H'_R = V(r) - V_{\text{Sph},R}(r).
\end{equation}

We are looking to minimize $\expval{H}$, which is
\begin{equation} \label{variationalHamiltonian}
\begin{split}
    \expval{H} &= \bra{\psi_{\text{Sph},R}} \left(H_{\text{Sph},R} + H'_R \right) \ket{\psi_{\text{Sph},R}} \\
    &= E_{\text{Sph},R} + \int \dd{r} \left( G_{\text{Sph},R}(r)^2 + F_{\text{Sph},R}(r)^2 \right) H'_R.
\end{split}
\end{equation}
For finding the minimum of this expression, any numerical opimization method, such as the golden-section search \cite{Kiefer1953} can be used.

If the value of $R$ for which $\expval{H}$ reaches its minimum is denoted as the effective radius $R_{\text{eff}}$, then the correction to the ground state energy is given approximately as
\begin{multline}
    \delta E_{\text{var}} = E_{\text{Sph},R_{\text{eff}}} - E_{\text{Sph},r_0} \\
    + \int \dd{r} \left( G_{\text{Sph},R_{\text{eff}}}(r)^2 + F_{\text{Sph},R_{\text{eff}}}(r)^2 \right) H'_{R_{\text{eff}}},
\end{multline}
and the $g$-factor correction is
\begin{equation}
    \delta g_{\text{var}} = \frac43 \pdv{\delta E_{\text{var}}}{M}.
\end{equation}
Again, the derivative of $E$ can be calculated from equation (\ref{EnergyMassDer}), but $r_0$ needs to be replaced with the effective radius $R_{\text{eff}}$.

\subsubsection{Numerically Solving the Dirac Equation}
The ground state wavefunctions and the corrections to the ground state energies can also be obtained from the numerical solution of the radial Dirac equation (\ref{radialDirac}). The wavefunctions are necessary for calculating the corrections to the $g$-factor using equation (\ref{gFactorIntegral}).
For the numerical results presented in this paper, we used the finite difference method to solve the radial Dirac equation (\ref{radialDirac}), and computed the $g$-factor corrections from
\begin{equation}
    \delta g_{\text{num}} = - \frac83 M \int \dd{r} r G(r) F(r) - g_{\text{Sph}}.
\end{equation}

\section{Alternative Charge Distribution Models}

\subsection{Two-Parameter Fermi Distribution}

\begin{table*}
\centering
\caption{\label{fermiTableE}Comparison of the corrections to the ground state energies for sphere and Fermi models, in units of the muon rest energy. $\Delta E_{\text{Sph}}$ is the leading order finite size correction (difference between the homogeneous sphere and point nucleus values), and $\delta E_{\text{Fermi}}$ are the Fermi distribution nucleus corrections to the sphere nucleus values. Values for the nuclear radii are taken from Ref. \cite{Angeli2013}. The parameter $a$ in Eq. (\ref{FermiChargeDist}) is taken to be $2.3\text{fm}/4\log{3}$ for all elements.}
\begin{ruledtabular}
\begin{tabular} {ccccccc}
& Z & $r_{\text{rms}}$ (fm)  & $\Delta E_{\text{Sph}}$ & $\delta E_{\text{Fermi}}^{\text{pert}}$ & $\delta E_{\text{Fermi}}^{\text{var}}$ & $\delta E_{\text{Fermi}}^{\text{num}}$ \\ \hline
$\prescript{12}{}{\text{C}}$  &  6  &  2.4702(22)  &   $3.8967(66) \times 10^{-6}$  &  $-2.3079(8) \times 10^{-8}$  &  $-2.3290(8) \times 10^{-8}$  &  $-2.3727(7) \times 10^{-8}$  \\
$\prescript{16}{}{\text{O}}$  &  8  &  2.6991(52)  &   $1.4057(50) \times 10^{-5}$  &  $-9.4133(63) \times 10^{-8}$  &  $-9.4931(60) \times 10^{-8}$  &  $-9.6493(57) \times 10^{-8}$  \\
$\prescript{20}{}{\text{Ne}}$  &  10  &  3.0055(21)  &   $4.0175(50) \times 10^{-5}$  &  $-2.7638(6) \times 10^{-7}$  &  $-2.7854(5) \times 10^{-7}$  &  $-2.8240(5) \times 10^{-7}$  \\
$\prescript{28}{}{\text{Si}}$  &  14  &  3.1224(24)  &   $1.5229(20) \times 10^{-4}$  &  $-1.2805(1) \times 10^{-6}$  &  $-1.2920(1) \times 10^{-6}$  &  $-1.3090(1) \times 10^{-6}$  \\
$\prescript{38}{}{\text{Ar}}$  &  18  &  3.4028(19)  &   $4.4039(38) \times 10^{-4}$  &  $-3.8294(1) \times 10^{-6}$  &  $-3.8660(1) \times 10^{-6}$  &  $-3.9098(2) \times 10^{-6}$  \\
$\prescript{40}{}{\text{Ca}}$  &  20  &  3.4776(19)  &   $6.6509(55) \times 10^{-4}$  &  $-5.9452(4) \times 10^{-6}$  &  $-6.0055(5) \times 10^{-6}$  &  $-6.0708(6) \times 10^{-6}$  \\
$\prescript{66}{}{\text{Zn}}$  &  30  &  3.9491(14)  &   $3.2385(14) \times 10^{-3}$  &  $-2.8141(5) \times 10^{-5}$  &  $-2.8487(6) \times 10^{-5}$  &  $-2.8730(6) \times 10^{-5}$  \\
$\prescript{86}{}{\text{Kr}}$  &  36  &  4.1835(21)  &   $6.3388(35) \times 10^{-3}$  &  $-5.2290(19) \times 10^{-5}$  &  $-5.2994(20) \times 10^{-5}$  &  $-5.3395(21) \times 10^{-5}$  \\
$\prescript{90}{}{\text{Zr}}$  &  40  &  4.2694(10)  &   $9.1096(22) \times 10^{-3}$  &  $-7.3846(15) \times 10^{-5}$  &  $-7.4906(15) \times 10^{-5}$  &  $-7.5446(16) \times 10^{-5}$  \\
$\prescript{120}{}{\text{Sn}}$  &  50  &  4.6519(21)  &   $1.9954(8) \times 10^{-2}$  &  $-1.3942(7) \times 10^{-4}$  &  $-1.4157(7) \times 10^{-4}$  &  $-1.4241(7) \times 10^{-4}$  \\
$\prescript{136}{}{\text{Xe}}$  &  54  &  4.7964(47)  &   $2.5930(21) \times 10^{-2}$  &  $-1.6995(20) \times 10^{-4}$  &  $-1.7262(21) \times 10^{-4}$  &  $-1.7357(21) \times 10^{-4}$  \\
$\prescript{142}{}{\text{Nd}}$  &  60  &  4.9123(25)  &   $3.6374(14) \times 10^{-2}$  &  $-2.2449(15) \times 10^{-4}$  &  $-2.2817(15) \times 10^{-4}$  &  $-2.2935(16) \times 10^{-4}$  \\
$\prescript{176}{}{\text{Yb}}$  &  70  &  5.3215(62)  &   $6.0941(44) \times 10^{-2}$  &  $-3.0765(53) \times 10^{-4}$  &  $-3.1265(55) \times 10^{-4}$  &  $-3.1397(55) \times 10^{-4}$  \\
$\prescript{185}{}{\text{Re}}$  &  75  &  5.3596(172)  &   $7.5168(139) \times 10^{-2}$  &  $-3.6513(181) \times 10^{-4}$  &  $-3.7125(187) \times 10^{-4}$  &  $-3.7277(189) \times 10^{-4}$  \\
$\prescript{208}{}{\text{Pb}}$  &  82  &  5.5012(13)  &   $9.9579(12) \times 10^{-2}$  &  $-4.4041(17) \times 10^{-4}$  &  $-4.4789(17) \times 10^{-4}$  &  $-4.4958(18) \times 10^{-4}$  \\
$\prescript{209}{}{\text{Bi}}$  &  83  &  5.5211(26)  &   $1.0346(2) \times 10^{-1}$  &  $-4.5137(35) \times 10^{-4}$  &  $-4.5904(36) \times 10^{-4}$  &  $-4.6076(36) \times 10^{-4}$  \\
$\prescript{212}{}{\text{Rn}}$  &  86  &  5.5915(176)  &   $1.1588(18) \times 10^{-1}$  &  $-4.8286(254) \times 10^{-4}$  &  $-4.9108(262) \times 10^{-4}$  &  $-4.9284(264) \times 10^{-4}$  \\
$\prescript{238}{}{\text{U}}$  &  92  &  5.8571(33)  &   $1.4530(4) \times 10^{-1}$  &  $-5.2566(51) \times 10^{-4}$  &  $-5.3428(53) \times 10^{-4}$  &  $-5.3598(53) \times 10^{-4}$  \\
\end{tabular}
\end{ruledtabular}
\end{table*}

\begin{table*}
\centering
\caption{\label{fermiTable}Comparison of the corrections to the $g$-factors for sphere and Fermi models. $\Delta g_{\text{Sph}}$ is the leading order finite size correction (difference between the homogeneous sphere and point nucleus values), and $\delta g_{\text{Fermi}}$ are the Fermi distribution nucleus corrections to the sphere nucleus values. Values for the nuclear radii are taken from Ref. \cite{Angeli2013}. The parameter $a$ in Eq. (\ref{FermiChargeDist}) is taken to be $2.3\text{fm}/4\log{3}$ for all elements.}
\begin{ruledtabular}
\begin{tabular} {ccccccc}
 & Z & $r_{\text{rms}}$ (fm)  & $\Delta g_{\text{Sph}}$ & $\delta g_{\text{Fermi}}^{\text{pert}}$ & $\delta g_{\text{Fermi}}^{\text{var}}$ & $\delta g_{\text{Fermi}}^{\text{num}}$ \\ \hline
$\prescript{12}{}{\text{C}}$  &  6  &  2.4702(22)  &   $1.5029(25) \times 10^{-5}$  &  $-1.1643(3) \times 10^{-7}$  &  $-1.1755(3) \times 10^{-7}$  &  $-1.1977(3) \times 10^{-7}$  \\
$\prescript{16}{}{\text{O}}$  &  8  &  2.6991(52)  &   $5.3243(183) \times 10^{-5}$  &  $-4.6250(24) \times 10^{-7}$  &  $-4.6676(23) \times 10^{-7}$  &  $-4.7447(21) \times 10^{-7}$  \\
$\prescript{20}{}{\text{Ne}}$  &  10  &  3.0055(21)  &   $1.4873(18) \times 10^{-4}$  &  $-1.3140(2) \times 10^{-6}$  &  $-1.3256(2) \times 10^{-6}$  &  $-1.3440(1) \times 10^{-6}$  \\
$\prescript{28}{}{\text{Si}}$  &  14  &  3.1224(24)  &   $5.4320(65) \times 10^{-4}$  &  $-5.7628(1) \times 10^{-6}$  &  $-5.8237(2) \times 10^{-6}$  &  $-5.8996(3) \times 10^{-6}$  \\
$\prescript{38}{}{\text{Ar}}$  &  18  &  3.4028(19)  &   $1.5004(12) \times 10^{-3}$  &  $-1.6118(2) \times 10^{-5}$  &  $-1.6308(2) \times 10^{-5}$  &  $-1.6490(3) \times 10^{-5}$  \\
$\prescript{40}{}{\text{Ca}}$  &  20  &  3.4776(19)  &   $2.2191(16) \times 10^{-3}$  &  $-2.4263(5) \times 10^{-5}$  &  $-2.4569(5) \times 10^{-5}$  &  $-2.4832(5) \times 10^{-5}$  \\
$\prescript{66}{}{\text{Zn}}$  &  30  &  3.9491(14)  &   $9.6827(35) \times 10^{-3}$  &  $-9.7629(29) \times 10^{-5}$  &  $-9.9198(30) \times 10^{-5}$  &  $-10.0011(31) \times 10^{-5}$  \\
$\prescript{86}{}{\text{Kr}}$  &  36  &  4.1835(21)  &   $1.7838(8) \times 10^{-2}$  &  $-1.6561(9) \times 10^{-4}$  &  $-1.6857(9) \times 10^{-4}$  &  $-1.6977(9) \times 10^{-4}$  \\
$\prescript{90}{}{\text{Zr}}$  &  40  &  4.2694(10)  &   $2.4763(5) \times 10^{-2}$  &  $-2.2189(6) \times 10^{-4}$  &  $-2.2614(7) \times 10^{-4}$  &  $-2.2766(7) \times 10^{-4}$  \\
$\prescript{120}{}{\text{Sn}}$  &  50  &  4.6519(21)  &   $4.9641(14) \times 10^{-2}$  &  $-3.6545(25) \times 10^{-4}$  &  $-3.7303(26) \times 10^{-4}$  &  $-3.7503(26) \times 10^{-4}$  \\
$\prescript{136}{}{\text{Xe}}$  &  54  &  4.7964(47)  &   $6.2469(35) \times 10^{-2}$  &  $-4.2363(66) \times 10^{-4}$  &  $-4.3259(69) \times 10^{-4}$  &  $-4.3471(69) \times 10^{-4}$  \\
$\prescript{142}{}{\text{Nd}}$  &  60  &  4.9123(25)  &   $8.4164(21) \times 10^{-2}$  &  $-5.2508(46) \times 10^{-4}$  &  $-5.3669(47) \times 10^{-4}$  &  $-5.3909(48) \times 10^{-4}$  \\
$\prescript{176}{}{\text{Yb}}$  &  70  &  5.3215(62)  &   $1.3139(6) \times 10^{-1}$  &  $-6.4182(141) \times 10^{-4}$  &  $-6.5585(147) \times 10^{-4}$  &  $-6.5813(148) \times 10^{-4}$  \\
$\prescript{185}{}{\text{Re}}$  &  75  &  5.3596(172)  &   $1.5801(18) \times 10^{-1}$  &  $-7.3148(459) \times 10^{-4}$  &  $-7.4795(477) \times 10^{-4}$  &  $-7.5041(481) \times 10^{-4}$  \\
$\prescript{208}{}{\text{Pb}}$  &  82  &  5.5012(13)  &   $2.0174(1) \times 10^{-1}$  &  $-8.3094(40) \times 10^{-4}$  &  $-8.4983(41) \times 10^{-4}$  &  $-8.5231(42) \times 10^{-4}$  \\
$\prescript{209}{}{\text{Bi}}$  &  83  &  5.5211(26)  &   $2.0856(3) \times 10^{-1}$  &  $-8.4463(81) \times 10^{-4}$  &  $-8.6385(84) \times 10^{-4}$  &  $-8.6632(85) \times 10^{-4}$  \\
$\prescript{212}{}{\text{Rn}}$  &  86  &  5.5915(176)  &   $2.3004(21) \times 10^{-1}$  &  $-8.8124(578) \times 10^{-4}$  &  $-9.0125(601) \times 10^{-4}$  &  $-9.0369(604) \times 10^{-4}$  \\
$\prescript{238}{}{\text{U}}$  &  92  &  5.8571(33)  &   $2.7897(4) \times 10^{-1}$  &  $-9.0619(110) \times 10^{-4}$  &  $-9.2593(114) \times 10^{-4}$  &  $-9.2804(114) \times 10^{-4}$  \\
\end{tabular}
\end{ruledtabular}
\end{table*}

A more realistic charge distribution model for the nucleus is the Fermi distribution,
\begin{equation} \label{FermiChargeDist}
    \rho_{\text{Fermi}}(\vb*{r}) = \frac{ZeN}{1+ \exponential(\frac{r-c}{a})},
\end{equation}
which depends on two adjustable parameters $a$ and $c$. Here, $N$ is a normalization constant that is chosen such that the normalization of the charge distribution is satisfied:
\begin{equation} \label{FermiNormalization}
    \int \dd[3]{r} \rho(\vb{r}) = Ze.
\end{equation}

The parameter $a$ is related to the skin thickness $t$ of the nucleus, which is the distance over which the charge density falls from 90\% to 10\%, as $t = 4a\log{3}$. For most nuclei, the skin thickness is approximately equal to 2.3fm \cite{Beier2000}, which was the value used for all of our calculations. With $a$ known, $c$ can be determined by requiring that the rms radius of the charge distribution (\ref{FermiChargeDist}) matches the rms radii of elements found in the literature, such as in \cite{Angeli2013}. The rms radius of a charge distribution is defined as
\begin{equation} \label{rmsRadiusGeneral}
    r_{\text{rms}} = \sqrt{\expval{r^2}} = \left( \frac{\int \dd[3]{r} r^2 \rho(\vb{r})}{\int \dd[3]{r} \rho(\vb{r})} \right)^{1/2},
\end{equation}
which, in the case of a spherically symmetric charge distribution, can be written as
\begin{equation} \label{rmsRadiusSpherical}
    r_{\text{rms}} = \sqrt{\expval{r^2}} = \left( \frac{\int_0^{\infty} \dd{r} r^4 \rho(r)}{\int_0^{\infty} \dd{r} r^2 \rho(r)} \right)^{1/2}.
\end{equation}

In the case of the Fermi distribution (\ref{FermiChargeDist}), the integrals can be evaluated analytically, and either approximative analytical formulas for $c$ and $N$, such as those given in \cite{Beier2000}, can be used, or, for more accurate results, $c$ and $N$ can be found by solving equations (\ref{rmsRadiusSpherical}) and (\ref{FermiNormalization}) numerically, which was the approach used for the results presented here.

The potential for the Fermi charge distribution (\ref{FermiChargeDist}) can also be calculated analytically, and is
\begin{multline} \label{FermiPotential}
    V_{\text{Fermi}}(r) = -4 \pi  a^2 N \alpha Z \left[\text{Li}_2\left(-e^{(c-r)/a}\right) \right. \\
    \left. +\frac{2 a}{r} \left[\text{Li}_3\left(-e^{(c-r)/a}\right)-\text{Li}_3\left(-e^{c/a}\right)\right]\right],
\end{multline}
where $\text{Li}_n(x)$ is the polylogarithm function \cite{Abramowitz1965}, which frequently appears in integrals involving the Fermi distribution.

An analytical solution of the Dirac equation with the Fermi distribution potential (\ref{FermiPotential}) is not known. Therefore, either approximation methods need to be used, or the Dirac equation needs to be solved numerically.

If the Fermi distribution potential is taken to be a perturbed sphere distribution potential, then $\delta V(r)$ in equation (\ref{pertPotential}) becomes
\begin{equation}
    \delta V(r) = V_{\text{Fermi}}(r) - V_{\text{Sph},r_0},
\end{equation}
and if the variational approximation is to be used, $H'_R$ in equation (\ref{varPotential}) becomes
\begin{equation}
    H'_R = V_{\text{Fermi}}(r) - V_{\text{Sph},R}.
\end{equation}
The results for both approximation methods, and for the numerical solution of the Dirac equation using finite difference method, is presented in Table \ref{fermiTableE} for the ground state energies, and in Table \ref{fermiTable} for the $g$-factors.

\subsection{Deformed Fermi Distribution and Nuclear Deformation Effect}

\begin{table}
    \centering
    \caption{\label{defParameters}Some isotopes and their nuclear deformation parameters. Values for the nuclear radii are taken from Ref. \cite{Angeli2013}, and nuclear deformation parameters are taken from \cite{Moller1995}, unless indicated otherwise. Only the elements with non-vanishing deformation parameters and even number of nucleons from Tables \ref{fermiTableE} and \ref{fermiTable} are considered.}
    \begin{ruledtabular}
    \begin{tabular}{ccccc}
         & Z & $r_{\text{rms}}$ (fm) & $\beta_2$ & $\beta_4$ \\ \hline
        $\prescript{12}{}{\text{C}}$  &  6  &  2.4702(22)  & 0.44\footnotemark[1] & 0.00\footnotemark[1] \\
        $\prescript{16}{}{\text{O}}$  &  8  &  2.6991(52)  &  0.021 & -0.108 \\
        $\prescript{20}{}{\text{Ne}}$  &  10  &  3.0055(21) & 0.335 & 0.428 \\
        $\prescript{28}{}{\text{Si}}$  &  14  &  3.1224(24)  & -0.478 & 0.250 \\
        $\prescript{66}{}{\text{Zn}}$  &  30  &  3.9491(14)  &  -0.215 & 0.005 \\
        $\prescript{86}{}{\text{Kr}}$  &  36  &  4.1835(21)  &  0.082 & 0.012 \\
        $\prescript{176}{}{\text{Yb}}$  &  70  &  5.3215(62)  &  0.278 & -0.071 \\
        $\prescript{208}{}{\text{Pb}}$  &  82  &  5.5012(13)  & 0.061 & 0.000 \\
        $\prescript{212}{}{\text{Rn}}$  &  86  &  5.5915(176)  &  0.000 & 0.008 \\
        $\prescript{238}{}{\text{U}}$  &  92  &  5.8571(33)  &  0.280\footnotemark[2] & 0.070\footnotemark[2]
    \end{tabular}
    \end{ruledtabular}
    \footnotetext[1]{Ref. \cite{Zatorski2012}}
\footnotetext[2]{Ref. \cite{Michel2019}}
\end{table}

\begin{table*}
\centering
\caption{\label{defFermiTableE}Comparison of the corrections to the ground state energies for the deformed Fermi model, in units of the muon rest energy. $\delta E_{\text{DefFermi}}$ are the differences between the deformed Fermi distribution nucleus results and the homogeneous sphere nucleus results, whereas $\delta E_{\text{ND}}$ are the differences between the deformed and undeformed Fermi distribution nucleus results. Nuclear radii and deformation parameters are listed in Table \ref{defParameters}. The parameter $a$ in Eq. (\ref{deformedFermi}) is taken to be $2.3\text{fm}/4\log{3}$ for all elements.}
\begin{ruledtabular}
    \begin{tabular} {ccccccc}
 & $-\delta E_{\text{DefFermi}}^{\text{pert}}$ & $-\delta E_{\text{DefFermi}}^{\text{var}}$ & $-\delta E_{\text{DefFermi}}^{\text{num}}$ & $-\delta E_{\text{ND}}^{\text{pert}}$ & $-\delta E_{\text{ND}}^{\text{var}}$ & $-\delta E_{\text{ND}}^{\text{num}}$\\ \hline
$\prescript{12}{6}{\text{C}}$  &   $2.354(1) \times 10^{-8}$  &  $2.377(1) \times 10^{-8}$  &  $2.423(1) \times 10^{-8}$  &  $4.663(38) \times 10^{-10}$  &  $4.756(39) \times 10^{-10}$  &  $4.983(40) \times 10^{-10}$  \\
$\prescript{16}{8}{\text{O}}$  &   $9.435(7) \times 10^{-8}$  &  $9.515(6) \times 10^{-8}$  &  $9.672(6) \times 10^{-8}$  &  $2.157(29) \times 10^{-10}$  &  $2.196(30) \times 10^{-10}$  &  $2.285(30) \times 10^{-10}$  \\
$\prescript{20}{10}{\text{Ne}}$  &   $3.271(3) \times 10^{-7}$  &  $3.301(3) \times 10^{-7}$  &  $3.357(3) \times 10^{-7}$  &  $5.069(20) \times 10^{-8}$  &  $5.157(20) \times 10^{-8}$  &  $5.326(20) \times 10^{-8}$  \\
$\prescript{28}{14}{\text{Si}}$  &   $1.4117(6) \times 10^{-6}$  &  $1.4258(6) \times 10^{-6}$  &  $1.4472(6) \times 10^{-6}$  &  $1.311(5) \times 10^{-7}$  &  $1.337(5) \times 10^{-7}$  &  $1.382(5) \times 10^{-7}$  \\
$\prescript{66}{30}{\text{Zn}}$  &   $2.9421(4) \times 10^{-5}$  &  $2.9801(4) \times 10^{-5}$  &  $3.0071(4) \times 10^{-5}$  &  $1.281(1) \times 10^{-6}$  &  $1.314(1) \times 10^{-6}$  &  $1.341(1) \times 10^{-6}$  \\
$\prescript{86}{36}{\text{Kr}}$  &   $5.285(2) \times 10^{-5}$  &  $5.357(2) \times 10^{-5}$  &  $5.398(2) \times 10^{-5}$  &  $5.584(7) \times 10^{-7}$  &  $5.740(7) \times 10^{-7}$  &  $5.835(7) \times 10^{-7}$  \\
$\prescript{90}{40}{\text{Zr}}$  &   $7.399(1) \times 10^{-5}$  &  $7.506(2) \times 10^{-5}$  &  $7.560(2) \times 10^{-5}$  &  $1.4721(8) \times 10^{-7}$  &  $1.5157(8) \times 10^{-7}$  &  $1.5400(8) \times 10^{-7}$  \\
$\prescript{176}{70}{\text{Yb}}$  &   $3.767(4) \times 10^{-4}$  &  $3.843(4) \times 10^{-4}$  &  $3.864(5) \times 10^{-4}$  &  $6.905(10) \times 10^{-5}$  &  $7.164(10) \times 10^{-5}$  &  $7.240(10) \times 10^{-5}$  \\
$\prescript{208}{82}{\text{Pb}}$  &   $4.459(2) \times 10^{-4}$  &  $4.536(2) \times 10^{-4}$  &  $4.554(2) \times 10^{-4}$  &  $5.535(1) \times 10^{-6}$  &  $5.729(1) \times 10^{-6}$  &  $5.776(1) \times 10^{-6}$  \\
$\prescript{212}{86}{\text{Rn}}$  &   $4.830(25) \times 10^{-4}$  &  $4.912(26) \times 10^{-4}$  &  $4.930(26) \times 10^{-4}$  &  $1.059(3) \times 10^{-7}$  &  $1.096(3) \times 10^{-7}$  &  $1.104(3) \times 10^{-7}$  \\
$\prescript{238}{92}{\text{U}}$  &   $7.380(4) \times 10^{-4}$  &  $7.551(4) \times 10^{-4}$  &  $7.585(4) \times 10^{-4}$  &  $2.124(1) \times 10^{-4}$  &  $2.209(1) \times 10^{-4}$  &  $2.225(1) \times 10^{-4}$  \\
\end{tabular}
\end{ruledtabular}
\end{table*}

\begin{table*}
\centering
\caption{\label{defFermiTable}Comparison of the corrections to the $g$-factors for the deformed Fermi model. $\delta g_{\text{DefFermi}}$ are the differences between the deformed Fermi distribution nucleus results and the homogeneous sphere nucleus results, whereas $\delta g_{\text{ND}}$ are the differences between the deformed and undeformed Fermi distribution nucleus results. Nuclear radii and deformation parameters are listed in Table \ref{defParameters}. The parameter $a$ in Eq. (\ref{deformedFermi}) is taken to be $2.3\text{fm}/4\log{3}$ for all elements.}
\begin{ruledtabular}
    \begin{tabular} {ccccccc}
& -$\delta g_{\text{DefFermi}}^{\text{pert}}$ & -$\delta g_{\text{DefFermi}}^{\text{var}}$ & -$\delta g_{\text{DefFermi}}^{\text{num}}$ & -$\delta g_{\text{ND}}^{\text{pert}}$ & -$\delta g_{\text{ND}}^{\text{var}}$ & -$\delta g_{\text{ND}}^{\text{num}}$ \\ \hline
$\prescript{12}{6}{\text{C}}$  &   $1.1878(5) \times 10^{-7}$  &  $1.1995(5) \times 10^{-7}$  &  $1.2229(5) \times 10^{-7}$  &  $2.355(19) \times 10^{-9}$  &  $2.405(19) \times 10^{-9}$  &  $2.520(20) \times 10^{-9}$  \\
$\prescript{16}{8}{\text{O}}$  &   $4.636(3) \times 10^{-7}$  &  $4.678(2) \times 10^{-7}$  &  $4.756(2) \times 10^{-7}$  &  $1.062(14) \times 10^{-9}$  &  $1.082(14) \times 10^{-9}$  &  $1.126(15) \times 10^{-9}$  \\
$\prescript{20}{10}{\text{Ne}}$  &   $1.555(1) \times 10^{-6}$  &  $1.571(1) \times 10^{-6}$  &  $1.598(1) \times 10^{-6}$  &  $2.411(9) \times 10^{-7}$  &  $2.458(9) \times 10^{-7}$  &  $2.539(9) \times 10^{-7}$  \\
$\prescript{28}{14}{\text{Si}}$  &   $6.355(2) \times 10^{-6}$  &  $6.429(2) \times 10^{-6}$  &  $6.525(2) \times 10^{-6}$  &  $5.919(21) \times 10^{-7}$  &  $6.057(22) \times 10^{-7}$  &  $6.259(23) \times 10^{-7}$  \\
$\prescript{66}{30}{\text{Zn}}$  &   $1.0210(2) \times 10^{-4}$  &  $1.0382(3) \times 10^{-4}$  &  $1.0472(3) \times 10^{-4}$  &  $4.469(4) \times 10^{-6}$  &  $4.622(4) \times 10^{-6}$  &  $4.713(4) \times 10^{-6}$  \\
$\prescript{86}{36}{\text{Kr}}$  &   $1.6739(9) \times 10^{-4}$  &  $1.7041(9) \times 10^{-4}$  &  $1.7164(9) \times 10^{-4}$  &  $1.777(2) \times 10^{-6}$  &  $1.842(2) \times 10^{-6}$  &  $1.871(2) \times 10^{-6}$  \\
$\prescript{90}{40}{\text{Zr}}$  &   $2.2233(6) \times 10^{-4}$  &  $2.2660(6) \times 10^{-4}$  &  $2.2813(7) \times 10^{-4}$  &  $4.449(2) \times 10^{-7}$  &  $4.625(2) \times 10^{-7}$  &  $4.695(2) \times 10^{-7}$  \\
$\prescript{176}{70}{\text{Yb}}$  &   $7.875(13) \times 10^{-4}$  &  $8.089(13) \times 10^{-4}$  &  $8.125(13) \times 10^{-4}$  &  $1.457(1) \times 10^{-4}$  &  $1.531(1) \times 10^{-4}$  &  $1.544(1) \times 10^{-4}$  \\
$\prescript{208}{82}{\text{Pb}}$  &   $8.415(4) \times 10^{-4}$  &  $8.609(4) \times 10^{-4}$  &  $8.634(4) \times 10^{-4}$  &  $1.0535(2) \times 10^{-5}$  &  $1.1032(2) \times 10^{-5}$  &  $1.1102(1) \times 10^{-5}$  \\
$\prescript{212}{86}{\text{Rn}}$  &   $8.814(58) \times 10^{-4}$  &  $9.015(60) \times 10^{-4}$  &  $9.039(60) \times 10^{-4}$  &  $1.952(3) \times 10^{-7}$  &  $2.045(5) \times 10^{-7}$  &  $2.056(3) \times 10^{-7}$  \\
$\prescript{238}{92}{\text{U}}$  &   $1.272(1) \times 10^{-3}$  &  $1.312(1) \times 10^{-3}$  &  $1.316(1) \times 10^{-3}$  &  $3.6606(8) \times 10^{-4}$  &  $3.8561(7) \times 10^{-4}$  &  $3.8767(7) \times 10^{-4}$  \\
    \end{tabular}
    \end{ruledtabular}
\end{table*}

Both the homogeneous sphere model, and the Fermi distribution model assume a spherically symmetric charge distribution for the nucleus, but the deformations of the nucleus can break its spherical symmetry. Therefore, it is more realistic to introduce a deformed Fermi distribution, in which the parameter $c$ has angular dependency. If the nucleus is still assumed to preserve its axial symmetry, then the dipole and the octupole deformations vanish, and in a first approximation, it is sufficient to consider quadrupole and hexadecapole deformations only \cite{Kozhedub2008, Zatorski2012, Michel2019}. This is achieved by replacing $c$ by $c_0(1 + \beta_2 Y_{20} + \beta_4 Y_{40})$, and writing the charge distribution as
\begin{equation} \label{deformedFermi}
    \rho_{\text{DefFermi}}(\vb{r}) = \frac{ZeN}{1+ \exponential(\frac{r-c_0(1 + \beta_2 Y_{20} + \beta_4 Y_{40})}{a})},
\end{equation}
where $Y_{20}$ and $Y_{40}$ are spherical harmonics, and $\beta_2$ and $\beta_4$ are quadrupole and octupole deformation parameters. The deformation parameters for some isotopes are listed in Table \ref{defParameters}.

The potential of the deformed Fermi distribution simplifies if the nucleus is spinless and is assumed to be in its rotational ground state, which means the nucleus has vanishing total angular momentum, which can only be satisfied if the nucleus has an even number of protons and neutrons. In the rotational ground state of the nucleus, the nucleus is found in all orientations with equal probability. If the nuclear and bound particle degrees of freedom are assumed to be independent, which is the approximation we consider in this work (this approximation is less applicable to muonic atoms as the excitation energies of the bound muon is similar in magnitude to the nuclear excitation energies, and the corrections due to the excitations of the nucleus are collectively denoted as the nuclear polarization effect, see the discussion in the last section), then the potential affecting the bound particle is the expectation value of the potential due to the nuclear charge distribution in the considered nuclear state \cite{Kozhedub2008}. In the case of the nuclear ground state, the expectation value averages out the angular dependence of the potential, and only the spherically symmetric component survives, in which case the potential for any charge distribution becomes
\begin{equation}
    V(r) = \frac{e}{4\pi} \int \dd[3]{r'} \frac{\rho(r')}{r_>},
\end{equation}
where $r_> = \max(r,r')$. 

In the case of the deformed Fermi distribution (\ref{deformedFermi}), the potential can be written as
\begin{widetext}
    \begin{equation}
    V_{\text{DefFermi}}(r) = -2 \pi  a^2 N \alpha Z \int_0^{\pi} \dd{\theta} \sin{\theta} \left[\text{Li}_2\left(-e^{(c(\theta)-r)/a}\right) + \frac{2 a}{r} \left[\text{Li}_3\left(-e^{(c(\theta)-r)/a}\right)-\text{Li}_3\left(-e^{c(\theta)/a}\right)\right]\right],
\end{equation}
\end{widetext}
where $c(\theta) = c_0(1 + \beta_2 Y_{20}(\theta) + \beta_4 Y_{40}(\theta))$. $c_0$ and $N$ can be found numerically using the formulas (\ref{rmsRadiusGeneral}) and (\ref{FermiNormalization}), with $a= \frac{2.3 \text{ fm}}{4\log{3}}$.

As in the Fermi distribution potential, the deformed Fermi distribution can also be considered a perturbation on top of the sphere nucleus potential with the same nuclear rms radius and charge, and perturbative and variational approximations can be applied, and the Dirac equation can also be solved numerically. The results for the ground state energies are presented in Table \ref{defFermiTableE}, and the results for the $g$-factor corrections in Table \ref{defFermiTable}. Nuclear deformation parameters are listed in Table~\ref{defParameters} and are usually taken from Ref.~\cite{Moller1995} unless stated otherwise.

It is important to notice that the corrections due to the difference between Fermi and Sphere charge distributions are larger than the uncertainties in the leading order finite nuclear size corrections, and for some elements large nuclear deformation parameters, even the nuclear deformation corrections, by which we refer to as the difference between the finite size corrections for deformed Fermi and un-deformed Fermi distributions, are larger than the uncertainties due to the rms radii, which is due to the nuclear effects being amplified in muonic atoms.

\begin{figure}
    \centering
    \includegraphics[width=\linewidth]{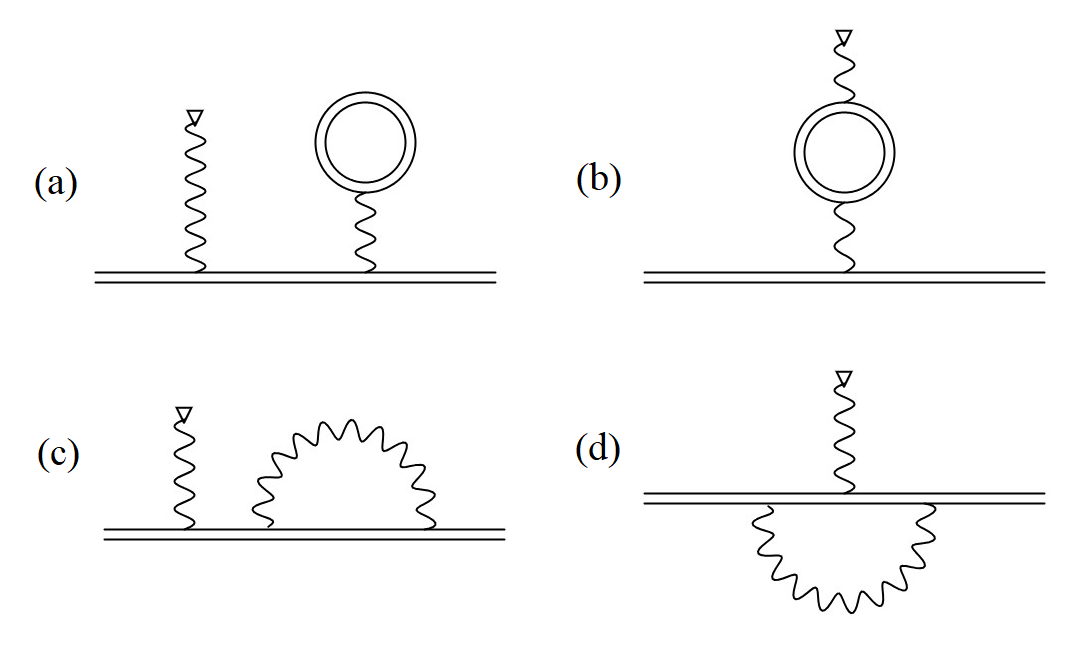}
    \caption{\label{OneLoopFeynmanDiagram}Feynman diagrams depicting the one-loop QED corrections to the $g$-factors: (a) electric-loop vacuum polarization correction, (b) magnetic-loop vacuum polarization correction, and (c, d) self-energy corrections.  Double lines represent bound-particle wave functions and propagators, the wave line ending with a triangle represents the constant external magnetic field. For the corrections to the energy levels, only the analogues of (a) and (c) appear, with the magnetic field line removed.}
\end{figure}

\section{Nuclear Size Corrections to the QED Effects}

So far, the corrections to the g-factors and energy levels have only been considered in the framework of relativistic quantum mechanics, but quantum electrodynamic (QED) effects also introduce corrections to the leading order finite size correction. At the one-loop level, these effects are the self-energy correction and the vacuum-polarization corrections. In the case of the $g$-factor, vacuum polarization manifests itself as electric-loop vacuum polarization and magnetic-loop vacuum polarization, where the magnetic-loop introduces corrections to the interaction with the external magnetic field, and the electric-loop introduces corrections to the interaction with the nuclear potential, and therefore, there is no analog of the magnetic loop in the Feynman diagrams for the QED corrections to the energy levels. The Feynman diagrams for the one-loop QED corrections to the $g$-factors are depicted in Figure~\ref{OneLoopFeynmanDiagram}.

\begin{figure}
    \centering
    \includegraphics[width=\linewidth]{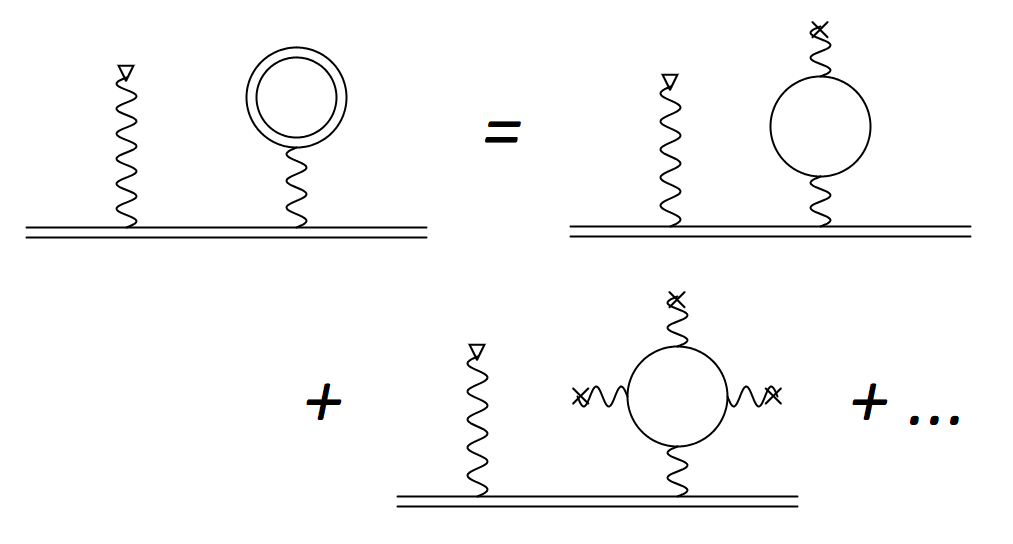}
    \caption{\label{VPdecomposition}Perturbative decomposition of the electric-loop vacuum polarization diagram into Uehling and Wichmann-Kroll diagrams. The first term on the right hand side is the Uehling diagram, and the remaining terms are collectively denoted as the Wichmann-Kroll terms. A double line represents the bound-particle propagator, and a single line represents the free particle propagator.}
\end{figure}

Electric-loop vacuum polarization describes the QED corrections to the electric field of the nucleus, and therefore it can be modelled as a correction to the nuclear potential. Since the bound-particle propagator is significantly more complicated than the free-particle propagator, the common approach in electric-loop vacuum polarization calculations is to perturbatively expand the loop in interactions with the nucleus, as depicted in terms of Feynman diagrams in Figure \ref{VPdecomposition}. To first order in the interaction with the nucleus, this correction potential is called the Uehling potential \cite{Uehling1935}. The Uehling potential changes depending on the particle that is pair-created, and we investigate the electronic, muonic, and hadronic vacuum polarization effects, as well as the dependence on the nuclear charge distribution. In the case of the electronic vacuum polarization effect in bound electrons, the leading Uehling contribution is found dominate over the higher-order electric loop corrections, the so-called Wichmann-Kroll term~\cite{Beier2000}. We therefore focus on the Uehling correction in this work.

Both the magnetic loop vacuum polarization correction and the self-energy correction require using the bound-muon propagator for extended nuclei for an accurate description~\cite{Yerokhin2013}. In this work, we limit ourselves to the electric-loop vacuum polarization corrections only, and leave the determination of finite nuclear size effects on the magnetic-loop vacuum polarization and self-energy corrections to the $g$-factors of bound muons for a future work.

\subsection{Electronic and Muonic Vacuum Polarization}

\begin{table*}
\centering
\caption{\label{eVPTableE}Electronic vacuum polarization corrections to the ground state energies in units of the muon rest energy. The results for both sphere-Uehling corrections ($\delta E_{e\text{VP,Sph}}$), and the Fermi-Uehling corrections ($\delta E_{e\text{VP,Fermi}}$) are presented and compared to the results in the case of a point-like nucleus ($\delta E_{e\text{VP,PN}}$). The nuclear radii and uncertainties are taken from Ref. \cite{Angeli2013}, same as in Table \ref{fermiTableE}.}
\begin{ruledtabular}
    \begin{tabular} {ccccccc}
 & $\delta E_{e\text{VP,PN}}^{\text{num}}$ & $\delta E_{e\text{VP,Sph}}^{\text{pert}}$ & $\delta E_{e\text{VP,Sph}}^{\text{var}}$ & $\delta E_{e\text{VP,Sph}}^{\text{num}}$ & $\delta E_{e\text{VP,Fermi}}^{\text{pert}}$ & $\delta E_{e\text{VP,Fermi}}^{\text{num}}$ \\ \hline
$\prescript{12}{6}{\text{C}}$  &   $-3.874 \times 10^{-6}$  &  $-3.8040(1) \times 10^{-6}$  &  $-3.8075(1) \times 10^{-6}$  &  $-3.8138(1) \times 10^{-6}$  &  $-3.8045(1) \times 10^{-6}$  &  $-3.8144(1) \times 10^{-6}$  \\
$\prescript{16}{8}{\text{O}}$  &   $-8.132 \times 10^{-6}$  &  $-7.8757(8) \times 10^{-6}$  &  $-7.8852(8) \times 10^{-6}$  &  $-7.8974(8) \times 10^{-6}$  &  $-7.8779(7) \times 10^{-6}$  &  $-7.8997(8) \times 10^{-6}$  \\
$\prescript{20}{10}{\text{Ne}}$  &   $-1.430 \times 10^{-5}$  &  $-1.3556(1) \times 10^{-5}$  &  $-1.3575(1) \times 10^{-5}$  &  $-1.3594(1) \times 10^{-5}$  &  $-1.3562(1) \times 10^{-5}$  &  $-1.3601(1) \times 10^{-5}$  \\
$\prescript{28}{14}{\text{Si}}$  &   $-3.298 \times 10^{-5}$  &  $-3.0040(3) \times 10^{-5}$  &  $-3.0092(3) \times 10^{-5}$  &  $-3.0128(3) \times 10^{-5}$  &  $-3.0071(3) \times 10^{-5}$  &  $-3.0160(4) \times 10^{-5}$  \\
$\prescript{38}{18}{\text{Ar}}$  &   $-6.102 \times 10^{-5}$  &  $-5.2362(7) \times 10^{-5}$  &  $-5.2461(7) \times 10^{-5}$  &  $-5.2513(7) \times 10^{-5}$  &  $-5.2454(6) \times 10^{-5}$  &  $-5.2605(7) \times 10^{-5}$  \\
$\prescript{40}{20}{\text{Ca}}$  &   $-7.883 \times 10^{-5}$  &  $-6.5658(9) \times 10^{-5}$  &  $-6.5784(9) \times 10^{-5}$  &  $-6.5844(9) \times 10^{-5}$  &  $-6.5799(9) \times 10^{-5}$  &  $-6.599(1) \times 10^{-5}$  \\
$\prescript{66}{30}{\text{Zn}}$  &   $-2.103 \times 10^{-4}$  &  $-1.4558(2) \times 10^{-4}$  &  $-1.4586(2) \times 10^{-4}$  &  $-1.4595(2) \times 10^{-4}$  &  $-1.4620(2) \times 10^{-4}$  &  $-1.4657(3) \times 10^{-4}$  \\
$\prescript{86}{36}{\text{Kr}}$  &   $-3.273 \times 10^{-4}$  &  $-2.0060(5) \times 10^{-4}$  &  $-2.0096(5) \times 10^{-4}$  &  $-2.0106(5) \times 10^{-4}$  &  $-2.0170(5) \times 10^{-4}$  &  $-2.0217(6) \times 10^{-4}$  \\
$\prescript{90}{40}{\text{Zr}}$  &   $-4.233 \times 10^{-4}$  &  $-2.4058(3) \times 10^{-4}$  &  $-2.4100(3) \times 10^{-4}$  &  $-2.4111(3) \times 10^{-4}$  &  $-2.4210(3) \times 10^{-4}$  &  $-2.4263(4) \times 10^{-4}$  \\
$\prescript{120}{50}{\text{Sn}}$  &   $-7.355 \times 10^{-4}$  &  $-3.356(1) \times 10^{-4}$  &  $-3.361(1) \times 10^{-4}$  &  $-3.362(1) \times 10^{-4}$  &  $-3.382(1) \times 10^{-4}$  &  $-3.388(1) \times 10^{-4}$  \\
$\prescript{136}{54}{\text{Xe}}$  &   $-8.934 \times 10^{-4}$  &  $-3.727(3) \times 10^{-4}$  &  $-3.733(3) \times 10^{-4}$  &  $-3.734(3) \times 10^{-4}$  &  $-3.758(3) \times 10^{-4}$  &  $-3.765(3) \times 10^{-4}$  \\
$\prescript{142}{60}{\text{Nd}}$  &   $-1.172 \times 10^{-3}$  &  $-4.334(2) \times 10^{-4}$  &  $-4.340(2) \times 10^{-4}$  &  $-4.341(2) \times 10^{-4}$  &  $-4.373(2) \times 10^{-4}$  &  $-4.381(2) \times 10^{-4}$  \\
$\prescript{176}{70}{\text{Yb}}$  &   $-1.768 \times 10^{-3}$  &  $-5.137(5) \times 10^{-4}$  &  $-5.144(5) \times 10^{-4}$  &  $-5.145(5) \times 10^{-4}$  &  $-5.188(5) \times 10^{-4}$  &  $-5.195(5) \times 10^{-4}$  \\
$\prescript{185}{75}{\text{Re}}$  &   $-2.144 \times 10^{-3}$  &  $-5.655(16) \times 10^{-4}$  &  $-5.661(16) \times 10^{-4}$  &  $-5.662(16) \times 10^{-4}$  &  $-5.714(17) \times 10^{-4}$  &  $-5.722(17) \times 10^{-4}$  \\
$\prescript{208}{82}{\text{Pb}}$  &   $-2.782 \times 10^{-3}$  &  $-6.284(1) \times 10^{-4}$  &  $-6.291(1) \times 10^{-4}$  &  $-6.292(1) \times 10^{-4}$  &  $-6.353(1) \times 10^{-4}$  &  $-6.361(1) \times 10^{-4}$  \\
$\prescript{209}{83}{\text{Bi}}$  &   $-2.886 \times 10^{-3}$  &  $-6.373(3) \times 10^{-4}$  &  $-6.380(3) \times 10^{-4}$  &  $-6.380(3) \times 10^{-4}$  &  $-6.443(3) \times 10^{-4}$  &  $-6.451(3) \times 10^{-4}$  \\
$\prescript{212}{86}{\text{Rn}}$  &   $-3.219 \times 10^{-3}$  &  $-6.622(20) \times 10^{-4}$  &  $-6.629(20) \times 10^{-4}$  &  $-6.630(20) \times 10^{-4}$  &  $-6.696(20) \times 10^{-4}$  &  $-6.704(20) \times 10^{-4}$  \\
$\prescript{238}{92}{\text{U}}$  &   $-4.003 \times 10^{-3}$  &  $-6.956(4) \times 10^{-4}$  &  $-6.963(4) \times 10^{-4}$  &  $-6.964(4) \times 10^{-4}$  &  $-7.034(4) \times 10^{-4}$  &  $-7.041(4) \times 10^{-4}$  \\
    \end{tabular}
    \end{ruledtabular}
\end{table*}

\begin{table*}
\centering
\caption{\label{eVPTableg}Electronic vacuum polarization corrections to the $g$-factors of particles in the ground state. The results for both sphere-Uehling corrections ($\delta g_{e\text{VP,Sph}}$), and the Fermi-Uehling corrections ($\delta g_{e\text{VP,Fermi}}$) are presented and compared to the results in the case of a point-like nucleus ($\delta g_{e\text{VP,PN}}$). The nuclear radii and uncertainties are taken from Ref. \cite{Angeli2013}, same as in Table \ref{fermiTable}.}
\begin{ruledtabular}
\begin{tabular} {ccccccc}
& $\delta g_{e\text{VP,PN}}^{\text{num}}$ & $\delta g_{e\text{VP,Sph}}^{\text{pert}}$ & $\delta g_{e\text{VP,Sph}}^{\text{var}}$ & $\delta g_{e\text{VP,Sph}}^{\text{num}}$ & $\delta g_{e\text{VP,Fermi}}^{\text{pert}}$ & $\delta g_{e\text{VP,Fermi}}^{\text{num}}$ \\ \hline
$\prescript{12}{6}{\text{C}}$  &   $-8.288 \times 10^{-6}$  &  $-8.0072(4) \times 10^{-6}$  &  $-8.0185(4) \times 10^{-6}$  &  $-8.0314(4) \times 10^{-6}$  &  $-8.0100(4) \times 10^{-6}$  &  $-8.0343(4) \times 10^{-6}$  \\
$\prescript{16}{8}{\text{O}}$  &   $-1.673 \times 10^{-5}$  &  $-1.5708(3) \times 10^{-5}$  &  $-1.5736(3) \times 10^{-5}$  &  $-1.5757(3) \times 10^{-5}$  &  $-1.5720(3) \times 10^{-5}$  &  $-1.5769(3) \times 10^{-5}$  \\
$\prescript{20}{10}{\text{Ne}}$  &   $-2.861 \times 10^{-5}$  &  $-2.5710(3) \times 10^{-5}$  &  $-2.5762(3) \times 10^{-5}$  &  $-2.5790(3) \times 10^{-5}$  &  $-2.5742(3) \times 10^{-5}$  &  $-2.5823(3) \times 10^{-5}$  \\
$\prescript{28}{14}{\text{Si}}$  &   $-6.353 \times 10^{-5}$  &  $-5.256(1) \times 10^{-5}$  &  $-5.267(1) \times 10^{-5}$  &  $-5.271(1) \times 10^{-5}$  &  $-5.270(1) \times 10^{-5}$  &  $-5.286(1) \times 10^{-5}$  \\
$\prescript{38}{18}{\text{Ar}}$  &   $-1.146 \times 10^{-4}$  &  $-8.406(2) \times 10^{-5}$  &  $-8.425(2) \times 10^{-5}$  &  $-8.429(2) \times 10^{-5}$  &  $-8.444(2) \times 10^{-5}$  &  $-8.468(2) \times 10^{-5}$  \\
$\prescript{40}{20}{\text{Ca}}$  &   $-1.466 \times 10^{-4}$  &  $-1.0130(3) \times 10^{-4}$  &  $-1.0152(3) \times 10^{-4}$  &  $-1.0156(3) \times 10^{-4}$  &  $-1.0187(3) \times 10^{-4}$  &  $-1.0214(3) \times 10^{-4}$  \\
$\prescript{66}{30}{\text{Zn}}$  &   $-3.775 \times 10^{-4}$  &  $-1.8264(5) \times 10^{-4}$  &  $-1.8294(5) \times 10^{-4}$  &  $-1.8295(5) \times 10^{-4}$  &  $-1.8468(5) \times 10^{-4}$  &  $-1.8501(6) \times 10^{-4}$  \\
$\prescript{86}{36}{\text{Kr}}$  &   $-5.791 \times 10^{-4}$  &  $-2.236(1) \times 10^{-4}$  &  $-2.239(1) \times 10^{-4}$  &  $-2.239(1) \times 10^{-4}$  &  $-2.269(1) \times 10^{-4}$  &  $-2.271(1) \times 10^{-4}$  \\
$\prescript{90}{40}{\text{Zr}}$  &   $-7.429 \times 10^{-4}$  &  $-2.5024(6) \times 10^{-4}$  &  $-2.5051(6) \times 10^{-4}$  &  $-2.5049(6) \times 10^{-4}$  &  $-2.5443(7) \times 10^{-4}$  &  $-2.5470(6) \times 10^{-4}$  \\
$\prescript{120}{50}{\text{Sn}}$  &   $-1.270 \times 10^{-3}$  &  $-2.909(2) \times 10^{-4}$  &  $-2.911(2) \times 10^{-4}$  &  $-2.910(2) \times 10^{-4}$  &  $-2.971(2) \times 10^{-4}$  &  $-2.972(2) \times 10^{-4}$  \\
$\prescript{136}{54}{\text{Xe}}$  &   $-1.534 \times 10^{-3}$  &  $-3.018(4) \times 10^{-4}$  &  $-3.020(4) \times 10^{-4}$  &  $-3.019(4) \times 10^{-4}$  &  $-3.087(4) \times 10^{-4}$  &  $-3.088(4) \times 10^{-4}$  \\
$\prescript{142}{60}{\text{Nd}}$  &   $-1.996 \times 10^{-3}$  &  $-3.218(2) \times 10^{-4}$  &  $-3.219(2) \times 10^{-4}$  &  $-3.218(2) \times 10^{-4}$  &  $-3.299(2) \times 10^{-4}$  &  $-3.299(2) \times 10^{-4}$  \\
$\prescript{176}{70}{\text{Yb}}$  &   $-2.979 \times 10^{-3}$  &  $-3.255(6) \times 10^{-4}$  &  $-3.255(6) \times 10^{-4}$  &  $-3.254(6) \times 10^{-4}$  &  $-3.345(6) \times 10^{-4}$  &  $-3.344(6) \times 10^{-4}$  \\
$\prescript{185}{75}{\text{Re}}$  &   $-3.593 \times 10^{-3}$  &  $-3.388(16) \times 10^{-4}$  &  $-3.387(16) \times 10^{-4}$  &  $-3.386(16) \times 10^{-4}$  &  $-3.487(17) \times 10^{-4}$  &  $-3.485(17) \times 10^{-4}$  \\
$\prescript{208}{82}{\text{Pb}}$  &   $-4.629 \times 10^{-3}$  &  $-3.461(1) \times 10^{-4}$  &  $-3.460(1) \times 10^{-4}$  &  $-3.459(1) \times 10^{-4}$  &  $-3.569(1) \times 10^{-4}$  &  $-3.567(1) \times 10^{-4}$  \\
$\prescript{209}{83}{\text{Bi}}$  &   $-4.797 \times 10^{-3}$  &  $-3.470(3) \times 10^{-4}$  &  $-3.468(3) \times 10^{-4}$  &  $-3.467(3) \times 10^{-4}$  &  $-3.578(3) \times 10^{-4}$  &  $-3.576(3) \times 10^{-4}$  \\
$\prescript{212}{86}{\text{Rn}}$  &   $-5.336 \times 10^{-3}$  &  $-3.481(17) \times 10^{-4}$  &  $-3.479(17) \times 10^{-4}$  &  $-3.478(17) \times 10^{-4}$  &  $-3.592(18) \times 10^{-4}$  &  $-3.589(18) \times 10^{-4}$  \\
$\prescript{238}{92}{\text{U}}$  &   $-6.596 \times 10^{-3}$  &  $-3.370(3) \times 10^{-4}$  &  $-3.368(3) \times 10^{-4}$  &  $-3.367(3) \times 10^{-4}$  &  $-3.479(3) \times 10^{-4}$  &  $-3.476(3) \times 10^{-4}$  \\
    \end{tabular}
    \end{ruledtabular}
\end{table*}

\begin{table*}
    \centering
    \caption{\label{muonicVP}Muonic vacuum polarization corrections to the ground state energies and $g$-factors. The energies are in units of muon rest mass. The results for both sphere-Uehling corrections and the Fermi-Uehling corrections are presented and compared to the results in the case of a point-like nucleus. Results obtained from the perturbative and variational approximations agree with the numerical results for up to 4 significant digits, therefore, only the numerical results are presented.}
    \begin{ruledtabular}
    \begin{tabular}{ccccccc}
    & $\delta E_{\mu\text{VP,PN}}$ & $\delta E_{\mu\text{VP,Sph}}$ & $\delta E_{\mu\text{VP,Fermi}}$ & $\delta g_{\mu\text{VP,PN}}$ & $\delta g_{\mu\text{VP,Sph}}$ & $\delta g_{\mu\text{VP,Fermi}}$ \\ \hline
$\prescript{12}{6}{\text{C}}$  &   $-2.177 \times 10^{-9}$  &  $-1.9142(3) \times 10^{-9}$  &  $-1.9206(3) \times 10^{-9}$  &  $-8.540 \times 10^{-9}$  &  $-7.179(1) \times 10^{-9}$  &  $-7.212(2) \times 10^{-9}$  \\
$\prescript{16}{8}{\text{O}}$  &   $-6.801 \times 10^{-9}$  &  $-5.584(3) \times 10^{-9}$  &  $-5.608(3) \times 10^{-9}$  &  $-2.654 \times 10^{-8}$  &  $-2.029(1) \times 10^{-8}$  &  $-2.042(1) \times 10^{-8}$  \\
$\prescript{20}{10}{\text{Ne}}$  &   $-1.643 \times 10^{-8}$  &  $-1.2328(3) \times 10^{-8}$  &  $-1.2391(3) \times 10^{-8}$  &  $-6.376 \times 10^{-8}$  &  $-4.305(1) \times 10^{-8}$  &  $-4.338(1) \times 10^{-8}$  \\
$\prescript{28}{14}{\text{Si}}$  &   $-6.205 \times 10^{-8}$  &  $-4.025(2) \times 10^{-8}$  &  $-4.056(2) \times 10^{-8}$  &  $-2.379 \times 10^{-7}$  &  $-1.3158(7) \times 10^{-7}$  &  $-1.3311(8) \times 10^{-7}$  \\
$\prescript{38}{18}{\text{Ar}}$  &   $-1.674 \times 10^{-7}$  &  $-8.969(3) \times 10^{-8}$  &  $-9.059(4) \times 10^{-8}$  &  $-6.338 \times 10^{-7}$  &  $-2.702(1) \times 10^{-7}$  &  $-2.744(1) \times 10^{-7}$  \\
$\prescript{40}{20}{\text{Ca}}$  &   $-2.539 \times 10^{-7}$  &  $-1.2443(5) \times 10^{-7}$  &  $-1.2583(5) \times 10^{-7}$  &  $-9.554 \times 10^{-7}$  &  $-3.610(2) \times 10^{-7}$  &  $-3.673(2) \times 10^{-7}$  \\
$\prescript{66}{30}{\text{Zn}}$  &   $-1.274 \times 10^{-6}$  &  $-3.759(2) \times 10^{-7}$  &  $-3.823(2) \times 10^{-7}$  &  $-4.642 \times 10^{-6}$  &  $-8.908(5) \times 10^{-7}$  &  $-9.157(5) \times 10^{-7}$  \\
$\prescript{86}{36}{\text{Kr}}$  &   $-2.657 \times 10^{-6}$  &  $-5.754(4) \times 10^{-7}$  &  $-5.870(4) \times 10^{-7}$  &  $-9.491 \times 10^{-6}$  &  $-1.214(1) \times 10^{-6}$  &  $-1.255(1) \times 10^{-6}$  \\
$\prescript{90}{40}{\text{Zr}}$  &   $-4.084 \times 10^{-6}$  &  $-7.344(3) \times 10^{-7}$  &  $-7.508(3) \times 10^{-7}$  &  $-1.439 \times 10^{-5}$  &  $-1.4490(7) \times 10^{-6}$  &  $-1.5038(7) \times 10^{-6}$  \\
$\prescript{120}{50}{\text{Sn}}$  &   $-1.035 \times 10^{-5}$  &  $-1.0985(8) \times 10^{-6}$  &  $-1.1269(9) \times 10^{-6}$  &  $-3.523 \times 10^{-5}$  &  $-1.811(2) \times 10^{-6}$  &  $-1.894(2) \times 10^{-6}$  \\
$\prescript{136}{54}{\text{Xe}}$  &   $-1.440 \times 10^{-5}$  &  $-1.238(2) \times 10^{-6}$  &  $-1.272(2) \times 10^{-6}$  &  $-4.828 \times 10^{-5}$  &  $-1.908(4) \times 10^{-6}$  &  $-2.001(5) \times 10^{-6}$  \\
$\prescript{142}{60}{\text{Nd}}$  &   $-2.285 \times 10^{-5}$  &  $-1.488(1) \times 10^{-6}$  &  $-1.532(2) \times 10^{-6}$  &  $-7.493 \times 10^{-5}$  &  $-2.105(3) \times 10^{-6}$  &  $-2.217(3) \times 10^{-6}$  \\
$\prescript{176}{70}{\text{Yb}}$  &   $-4.634 \times 10^{-5}$  &  $-1.739(4) \times 10^{-6}$  &  $-1.793(4) \times 10^{-6}$  &  $-1.460 \times 10^{-4}$  &  $-2.096(7) \times 10^{-6}$  &  $-2.217(7) \times 10^{-6}$  \\
$\prescript{185}{75}{\text{Re}}$  &   $-6.467 \times 10^{-5}$  &  $-1.957(13) \times 10^{-6}$  &  $-2.020(14) \times 10^{-6}$  &  $-1.994 \times 10^{-4}$  &  $-2.229(20) \times 10^{-6}$  &  $-2.365(21) \times 10^{-6}$  \\
$\prescript{208}{82}{\text{Pb}}$  &   $-1.017 \times 10^{-4}$  &  $-2.186(1) \times 10^{-6}$  &  $-2.260(1) \times 10^{-6}$  &  $-3.039 \times 10^{-4}$  &  $-2.284(2) \times 10^{-6}$  &  $-2.432(2) \times 10^{-6}$  \\
$\prescript{209}{83}{\text{Bi}}$  &   $-1.084 \times 10^{-4}$  &  $-2.217(2) \times 10^{-6}$  &  $-2.292(2) \times 10^{-6}$  &  $-3.224 \times 10^{-4}$  &  $-2.289(3) \times 10^{-6}$  &  $-2.439(3) \times 10^{-6}$  \\
$\prescript{212}{86}{\text{Rn}}$  &   $-1.313 \times 10^{-4}$  &  $-2.299(16) \times 10^{-6}$  &  $-2.378(17) \times 10^{-6}$  &  $-3.847 \times 10^{-4}$  &  $-2.288(21) \times 10^{-6}$  &  $-2.441(23) \times 10^{-6}$  \\
$\prescript{238}{92}{\text{U}}$  &   $-1.924 \times 10^{-4}$  &  $-2.333(3) \times 10^{-6}$  &  $-2.412(3) \times 10^{-6}$  &  $-5.465 \times 10^{-4}$  &  $-2.130(4) \times 10^{-6}$  &  $-2.275(4) \times 10^{-6}$  \\
    \end{tabular}
    \end{ruledtabular}
\end{table*}

Since electrons and muons are particles with identical properties except for their mass, electronic and muonic Uehling potentials are given by the same expression. The Uehling potential for the case of leptonic vacuum polarization and extended nucleus can be expressed as~\cite{QFT}
\begin{equation}
V_{\mathrm{Ueh}} = \frac{2}{4 \pi^2} \int_{2 m_l}^{\infty} \mathrm{d} q \, \rho_{\mathrm{int}} (r,q) \Im \Pi (q^2).
\end{equation}
The imaginary part of the leptonic vacuum polarization function is
\begin{equation}
\Im \Pi (q^2) = \frac{\alpha}{3} \sqrt{1 - \frac{4 m_l^2}{q^2}} \left( 1 + \frac{2 m_l^2}{q^2} \right).
\end{equation}
where $m_l$ is the mass of the virtual lepton that is pair-created, being either $m_e$ or $m_{\mu}$. The integral over the nuclear charge distribution
\begin{equation}
\rho_{\mathrm{int}} (r,q) = -\frac{2 \pi e}{q^2 r} \int_0^\infty \mathrm{d} r' r' \rho (r') \left( e^{-q \vert r - r' \vert} - e^{-q (r + r')} \right)
\end{equation}
was performed analytically for the sphere distribution, and numerically for the Fermi distribution.

Depending on whether the homogeneous sphere model, or the Fermi distribution model is used, the total potential becomes either $V(r) = V_{\text{Sph}}(r) + V_{\text{Ueh-Sph},m}(r)$, or $V(r) = V_{\text{Fermi}}(r) + V_{\text{Ueh-Fermi},m}(r)$. In the case of the homogeneous sphere model, perturbative and variational approximations can be applied, or the Dirac equation can be solved numerically. In the case of the Fermi distribution model, the Fermi-Uehling potential can be considered as a perturbation on top of the Fermi nucleus potential (as in Eq. (\ref{pertPotential}), but with $V_{\text{Sph}}$ replaced by $V_{\text{Fermi}}$), and perturbation theory using the numerical wavefunctions can be applied, or the Dirac equation can be solved numerically for the total potential. For the perturbation theory calculation of the Fermi-Uehling contribution to the $g$-factor, Eq. (\ref{perturbativeGFactor}) needs to be used, in which the mass derivatives of the wavefunctions appear. If the Fermi-distribution nucleus wavefunctions are found numerically, their mass derivatives can be obtained by solving the mass derivative of the Dirac equation
\begin{equation}
    \left( H - E \right) \pdv{M}\ket{\psi(\vb{r})} = \left(\pdv{E}{M} - \beta \right) \ket{\psi(\vb{r})},
\end{equation}
or in terms of the radial wavefunctions,
\begin{multline}
    \begin{pmatrix}
        M + V(r) - E & -\dv{r} + \frac{\kappa}{r} \\ \dv{r} + \frac{\kappa}{r} & - M + V(r) - E
    \end{pmatrix} \begin{pmatrix}
        G'(r) \\ F'(r)
    \end{pmatrix} = \\ \begin{pmatrix}
        (E' - 1) G(r) \\ (E' + 1) F(r)
    \end{pmatrix}
\end{multline}
and requiring that $ \bra{\psi(\vb{r})}\pdv{M}\ket{\psi(\vb{r})} = 0$.

Electronic vacuum polarization results for the ground state energies are presented in Table \ref{eVPTableE}, and for the $g$-factors in Table \ref{eVPTableg}. Muonic vacuum polarization results are presented in Table \ref{muonicVP}.

\begin{table*}
    \centering
    \caption{\label{hadrVPTable}Hadronic vacuum polarization corrections to the ground state energies and $g$-factors. The energies are in units of muon rest mass. The results for both sphere-Uehling corrections and the Fermi-Uehling corrections are presented and compared to the results in the case of a point-like nucleus. Results obtained from the perturbative and variational approximations agree with the numerical results for up to 4 significant digits, therefore, only the numerical results are presented.}
    \begin{ruledtabular}
    \begin{tabular}{ccccccc}
& $\delta E_{h\text{VP,PN}}$ & $\delta E_{h\text{VP,Sph}}$ & $\delta E_{h\text{VP,Fermi}}$ & $\delta g_{h\text{VP,PN}}$ & $\delta g_{h\text{VP,Sph}}$ & $\delta g_{h\text{VP,Fermi}}$ \\ \hline
$\prescript{12}{6}{\text{C}}$  &   $-1.492 \times 10^{-9}$  &  $-1.2838(2) \times 10^{-9}$  &  $-1.2885(4) \times 10^{-9}$  &  $-5.904 \times 10^{-9}$  &  $-4.829(2) \times 10^{-9}$  &  $-4.854(1) \times 10^{-9}$  \\
$\prescript{16}{8}{\text{O}}$  &   $-4.710 \times 10^{-9}$  &  $-3.752(2) \times 10^{-9}$  &  $-3.770(2) \times 10^{-9}$  &  $-1.860 \times 10^{-8}$  &  $-1.368(1) \times 10^{-8}$  &  $-1.378(1) \times 10^{-8}$  \\
$\prescript{20}{10}{\text{Ne}}$  &   $-1.150 \times 10^{-8}$  &  $-8.293(2) \times 10^{-9}$  &  $-8.339(2) \times 10^{-9}$  &  $-4.528 \times 10^{-8}$  &  $-2.908(1) \times 10^{-8}$  &  $-2.932(1) \times 10^{-8}$  \\
$\prescript{28}{14}{\text{Si}}$  &   $-4.434 \times 10^{-8}$  &  $-2.717(1) \times 10^{-8}$  &  $-2.738(3) \times 10^{-8}$  &  $-1.735 \times 10^{-7}$  &  $-8.931(5) \times 10^{-8}$  &  $-9.037(10) \times 10^{-8}$  \\
$\prescript{38}{18}{\text{Ar}}$  &   $-1.222 \times 10^{-7}$  &  $-6.068(2) \times 10^{-8}$  &  $-6.131(3) \times 10^{-8}$  &  $-4.748 \times 10^{-7}$  &  $-1.839(1) \times 10^{-7}$  &  $-1.869(1) \times 10^{-7}$  \\
$\prescript{40}{20}{\text{Ca}}$  &   $-1.875 \times 10^{-7}$  &  $-8.428(4) \times 10^{-8}$  &  $-8.527(5) \times 10^{-8}$  &  $-7.253 \times 10^{-7}$  &  $-2.461(1) \times 10^{-7}$  &  $-2.507(2) \times 10^{-7}$  \\
$\prescript{66}{30}{\text{Zn}}$  &   $-9.955 \times 10^{-7}$  &  $-2.554(1) \times 10^{-7}$  &  $-2.600(1) \times 10^{-7}$  &  $-3.771 \times 10^{-6}$  &  $-6.102(3) \times 10^{-7}$  &  $-6.279(4) \times 10^{-7}$  \\
$\prescript{86}{36}{\text{Kr}}$  &   $-2.152 \times 10^{-6}$  &  $-3.914(3) \times 10^{-7}$  &  $-3.996(3) \times 10^{-7}$  &  $-8.038 \times 10^{-6}$  &  $-8.333(8) \times 10^{-7}$  &  $-8.625(8) \times 10^{-7}$  \\
$\prescript{90}{40}{\text{Zr}}$  &   $-3.390 \times 10^{-6}$  &  $-5.000(2) \times 10^{-7}$  &  $-5.116(2) \times 10^{-7}$  &  $-1.254 \times 10^{-5}$  &  $-9.955(5) \times 10^{-7}$  &  $-1.0346(5) \times 10^{-6}$  \\
$\prescript{120}{50}{\text{Sn}}$  &   $-9.162 \times 10^{-6}$  &  $-7.480(6) \times 10^{-7}$  &  $-7.681(6) \times 10^{-7}$  &  $-3.296 \times 10^{-5}$  &  $-1.245(1) \times 10^{-6}$  &  $-1.304(1) \times 10^{-6}$  \\
$\prescript{136}{54}{\text{Xe}}$  &   $-1.308 \times 10^{-5}$  &  $-8.431(15) \times 10^{-7}$  &  $-8.667(16) \times 10^{-7}$  &  $-4.652 \times 10^{-5}$  &  $-1.311(3) \times 10^{-6}$  &  $-1.377(3) \times 10^{-6}$  \\
$\prescript{142}{60}{\text{Nd}}$  &   $-2.165 \times 10^{-5}$  &  $-1.014(1) \times 10^{-6}$  &  $-1.044(1) \times 10^{-6}$  &  $-7.548 \times 10^{-5}$  &  $-1.447(2) \times 10^{-6}$  &  $-1.526(2) \times 10^{-6}$  \\
$\prescript{176}{70}{\text{Yb}}$  &   $-4.722 \times 10^{-5}$  &  $-1.183(3) \times 10^{-6}$  &  $-1.220(3) \times 10^{-6}$  &  $-1.589 \times 10^{-4}$  &  $-1.437(5) \times 10^{-6}$  &  $-1.523(5) \times 10^{-6}$  \\
$\prescript{185}{75}{\text{Re}}$  &   $-6.849 \times 10^{-5}$  &  $-1.331(9) \times 10^{-6}$  &  $-1.376(10) \times 10^{-6}$  &  $-2.259 \times 10^{-4}$  &  $-1.529(14) \times 10^{-6}$  &  $-1.625(15) \times 10^{-6}$  \\
$\prescript{208}{82}{\text{Pb}}$  &   $-1.141 \times 10^{-4}$  &  $-1.4865(8) \times 10^{-6}$  &  $-1.5381(8) \times 10^{-6}$  &  $-3.650 \times 10^{-4}$  &  $-1.565(1) \times 10^{-6}$  &  $-1.671(1) \times 10^{-6}$  \\
$\prescript{209}{83}{\text{Bi}}$  &   $-1.226 \times 10^{-4}$  &  $-1.508(2) \times 10^{-6}$  &  $-1.560(2) \times 10^{-6}$  &  $-3.906 \times 10^{-4}$  &  $-1.568(2) \times 10^{-6}$  &  $-1.675(2) \times 10^{-6}$  \\
$\prescript{212}{86}{\text{Rn}}$  &   $-1.524 \times 10^{-4}$  &  $-1.563(11) \times 10^{-6}$  &  $-1.618(12) \times 10^{-6}$  &  $-4.785 \times 10^{-4}$  &  $-1.567(14) \times 10^{-6}$  &  $-1.676(16) \times 10^{-6}$  \\
$\prescript{238}{92}{\text{U}}$  &   $-2.359 \times 10^{-4}$  &  $-1.584(2) \times 10^{-6}$  &  $-1.639(2) \times 10^{-6}$  &  $-7.182 \times 10^{-4}$  &  $-1.455(2) \times 10^{-6}$  &  $-1.558(3) \times 10^{-6}$  \\
    \end{tabular}
    \end{ruledtabular}
\end{table*}

\subsection{Hadronic Vacuum Polarization}

Unlike leptons, hadrons are composite particles that interact and are bound together via the strong interaction, which is decribed by the theory of quantum chromodynamics (QCD). In QCD, unlike in QED, the perturbative calculation methods fail due to the relatively strong coupling of the quark and gluon fields \cite{Jegerlehner2002}, which renders the calculation of the interaction of virtual hadronic particles a difficult task, and the hadronic vacuum polarization function appearing in the formula for the Uehling potential is difficult to calculate from theory alone. Instead, a common method in the calculations of the hadronic vacuum polarization is to construct a polarization function from experimental data \cite{Burkhardt1989}. Using this polarization function, it is possible to calculate the Uehling potential, as was done in \cite{Breidenbach2022,Dizer2023}.

The Uehling potential for an arbitrary polarization function $\Pi$ and a spherically symmetric charge distribution can be expressed as \cite{Greiner2009}
\begin{equation} \label{Vuehhadr}
    V_{\text{Ueh}}(r) = -\frac{e}{2\pi^2} \int_0^{\infty} \dd{q} j_0(q r) \Tilde{\rho}(q) \Re[\Pi^R(-q^2)],
\end{equation}
where $j_0(x)$ is the spherical Bessel function of the first kind, $\Tilde{\rho}(q)$ is the Fourier transform of the charge distribution $\rho(r)$, and $\Pi^R(-q^2)$ is the regular part of the polarization function.

In the case of the hadronic vacuum polarization, the polarization function is written as \cite{Burkhardt2004}
\begin{equation} \label{hadronicPolarizationFunction}
    \Re[\Pi^R(q^2)] = A_i + B_i \log(1 + C_i \abs{q^2}),
\end{equation}
where $A_i$, $B_i$, and $C_i$ have different values depending on the range of $q$, as given in Table \ref{hadrPiRanges}. 

\begin{table}
\caption{\label{hadrPiRanges}Piecewise parametrization of the hadronic polarization function in Eq. (\ref{hadronicPolarizationFunction}), as given in \cite{Burkhardt2004, Breidenbach2022} with the mass of the Z boson $m_Z=91.1876 \text{ GeV}$. Adapted from Ref. \cite{Breidenbach2022}.}
    \centering
    \begin{ruledtabular}
       \begin{tabular}{cccc}
    $q$ (GeV) & $A_i$ & $B_i$ & $C_i \,(\text{GeV}^{-2})$ \\ \hline
    0.0$-$0.7 & 0.0 & 0.0023092 & 3.9925370 \\
    0.7$-$2.0 & 0.0 & 0.0022333 & 4.2191779 \\
    2.0$-$4.0 & 0.0 & 0.0024402 & 3.2496684 \\
    4.0$-$10.0 & 0.0 & 0.0027340 & 2.0995092 \\
    10.0$-$$m_Z$ & 0.0010485 & 0.0029431 & 1.0 \\
    $m_Z$$-$$10^4$ & 0.0012234 & 0.0029237 & 1.0 \\
    $10^4$$-$$10^5$ & 0.0016894 & 0.0028984 & 1.0
    \end{tabular} 
    \end{ruledtabular}
\end{table}

In our investigation, we obtained results both by numerical integration of the expression in Eq. (\ref{Vuehhadr}) using the piecewise defined polarization function in Eq. (\ref{hadronicPolarizationFunction}) and Table \ref{hadrPiRanges}, and by using the analytical formula for the hadronic Uehling potential for homogeneous sphere charge distribution nucleus, derived in \cite{Breidenbach2022} with the approximation that the low energy region of $q$ in Eq. (\ref{hadronicPolarizationFunction}) can be extended to infinity, and the values $A =0$, $B = 0.0023092$, and $C = 3.9925370 \text{ GeV}^{-2}$ can be used for the entire range of $q$. This approximation works better for the extended nucleus Uehling potential than the point-like nucleus Uehling potential, as the Fourier transform $\Tilde{\rho}(q)$ of the nuclear charge distribution appearing in Eq. (\ref{Vuehhadr}) is
\begin{equation}
    \Tilde{\rho}(q) = 3 Z e \frac{j_1(q r_0)}{q r_0},
\end{equation}
which is a function approaching $0$ as $q \rightarrow \infty$ for the extended nucleus, but becomes $\Tilde{\rho}(q) = Z e$ for the point-like nucleus ($r_0 \rightarrow 0$), meaning that the contributions of the large $q$ regions in Eq. (\ref{Vuehhadr}) is greater for the point-like nucleus than the extended nucleus.
We found that the results obtained from both methods agree in the first three decimal digits for the extended nucleus case, and the difference between both methods is smaller than the rms radius uncertainties in the hadronic vacuum polarization corrections. In Table \ref{hadrVPTable}, we present the results that were obtained numerically with the piecewise defined polarization function in Eq. (\ref{hadronicPolarizationFunction}) and Table \ref{hadrPiRanges}.

\section{Discussion of Results}

\begin{figure*}
    \centering
    \includegraphics[width=\textwidth]{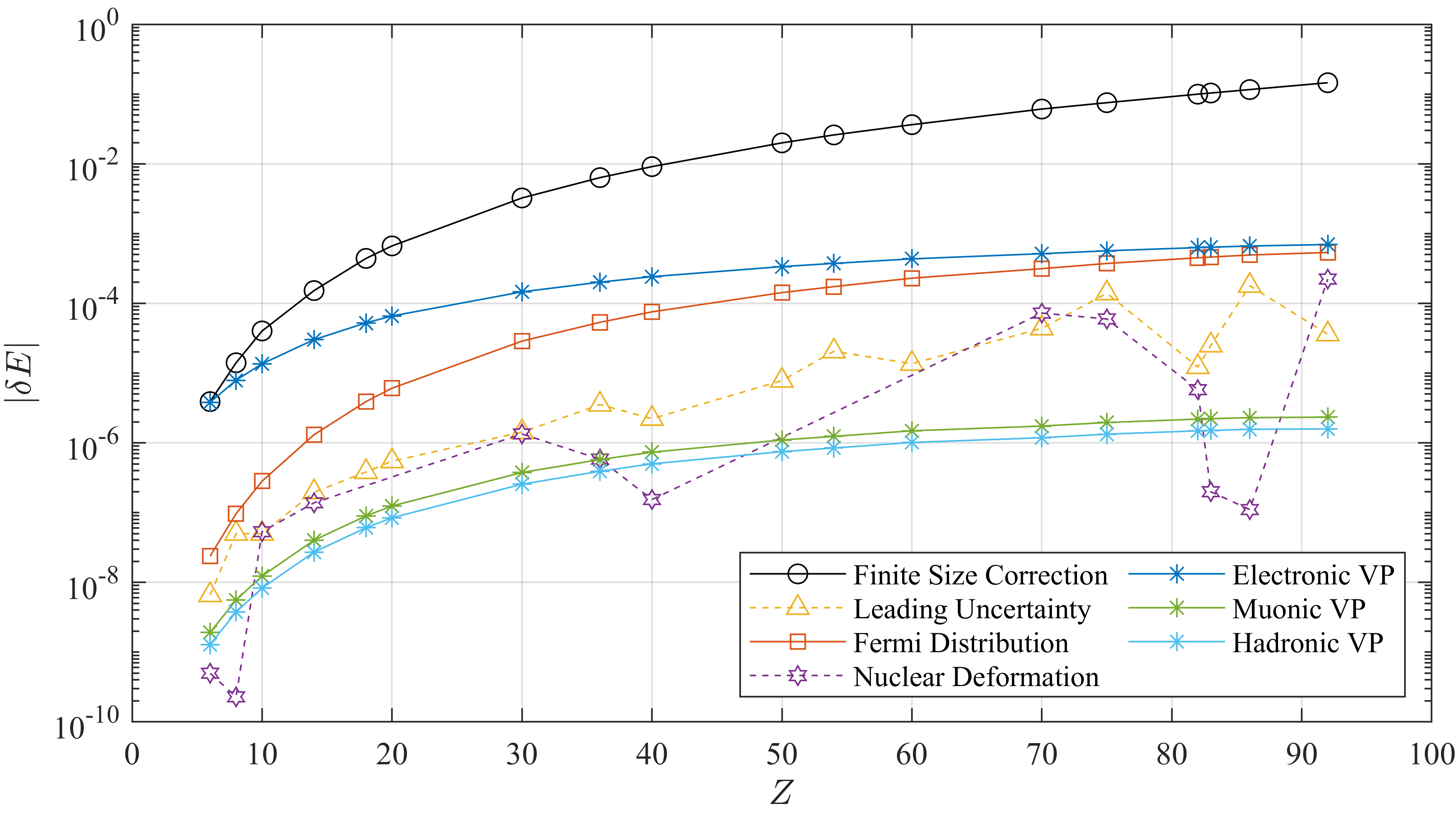}
    \caption{\label{delta_E_figure}Comparison of the magnitudes of various contributions to the ground state energies of bound muons for several atoms. The plotted vacuum polarization corrections are for the spherical nucleus case. The leading uncertainty is the uncertainty in the leading finite size corrections due to the rms radius uncertainties.}
\end{figure*}

\begin{figure*}
    \centering
    \includegraphics[width=\textwidth]{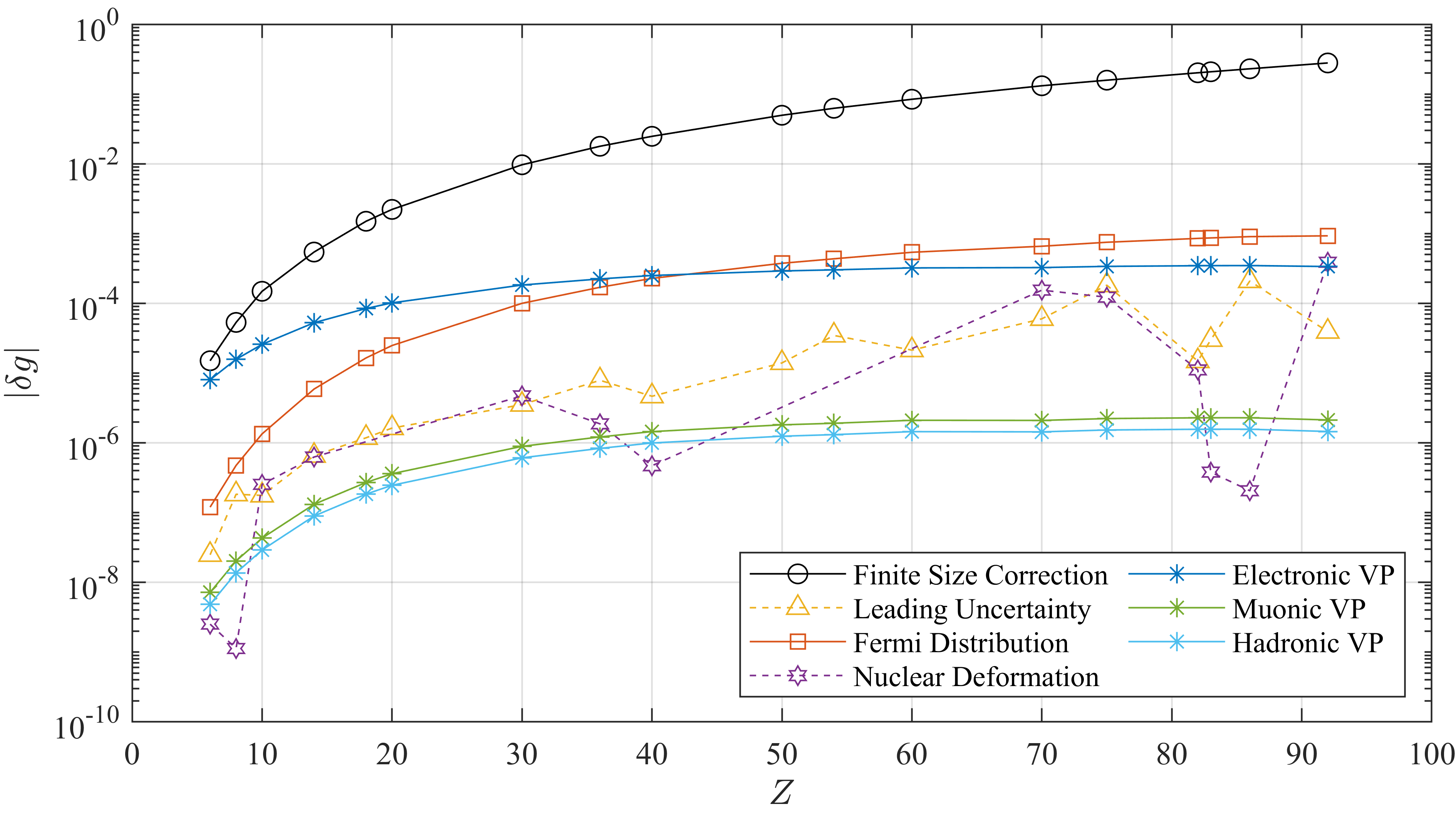}
    \caption{\label{delta_g_figure}Comparison of the magnitudes of various contributions to the $g$-factors of bound muons for several atoms. The plotted vacuum polarization corrections are for the spherical nucleus case. The leading uncertainty is the uncertainty in the leading finite size corrections due to the rms radius uncertainties.}
\end{figure*}

An overview of the corrections to the bound muon ground state energies and $g$-factors that were discussed so far are depicted in Figures \ref{delta_E_figure} and \ref{delta_g_figure}, together with the uncertainty in the leading finite size corrections due to the known rms radius uncertainties.

In comparison to electronic atoms, nuclear effects in muonic atoms are amplified due to the larger mass of the muon, which manifests itself clearly in the corrections to the ground state energies and $g$-factors. The nuclear charge distribution model dependence is amplified, and is larger than the leading uncertainties for all elements, and for nuclei with large octupole and quadrupole deformations, even the nuclear deformation corrections can be larger than the uncertainties due to the rms radii, which is importantly the case for uranium. This is different from the case of the bound-electron $g$-factor in light ions, where the nuclear model uncertainty is much smaller than the uncertainty due to the rms radii~\cite{Sturm11,Zatorski2012,Yerokhin2013}.

Unlike the nuclear model dependence, the finite size of the nucleus causes the vacuum polarization corrections to be suppressed in comparison to the vacuum polarization corrections in the case of a point-like nucleus, and the suppression becomes more pronounced for heavier elements. This can be conceptually understood as follows: for a finite-size nucleus model, the potential is the same as the Coulomb potential outside the nucleus, but is smaller in magnitude inside the nucleus. Therefore, virtual particles that are created inside the nucleus is under the effect of a weaker potential, which results in a weaker vacuum polarization and shielding of the nuclear charge. This suppression of the electric-loop vacuum polarization corrections become more evident for the $g$-factors, and for heavier elements, difference of leading finite size corrections for different nuclear models due to the different nuclear charge distribution models become even greater than the electronic vacuum polarization corrections, as can be seen in Figure \ref{delta_g_figure}.

In the case of the free muon $g$-factor, electronic vacuum polarization is found to be strongly enhanced compared to electronic vacuum polarization in the free-electron $g$-factor~\cite{CODATA2018}, since vacuum polarization effects depend only on the mass ratio of the reference lepton and the lepton in the vacuum polarization loop. Similarly, we find electronic vacuum polarization in the bound-muon $g$-factor to be enhanced compared to the bound-electron $g$-factor (see e.g. results in Ref.~\cite{Beier2000}.) However, for heavy muonic atoms, for $Z>86$, we find the suppression of the vacuum polarization effect due to the finite nuclear size to be stronger than the expected enhancement due to the lepton mass ratio. Therefore, surprisingly, the absolute impact of electronic vacuum polarization is smaller in the heaviest muonic atoms than in the corresponding electronic H-like ions.

The muonic vacuum polarization contribution to the $g$-factor in muonic atoms is found to be smaller than electronic vacuum polarization in electronic ions, due to the stronger suppression due to the finite nuclear size. (In point-nucleus calculations, electronic vacuum polarization in electronic ions and muonic vacuum polarization in muonic ions are identical.)

The suppression in the muonic and hadronic vacuum polarization corrections are even larger than the suppression to the electronic vacuum polarization corrections: the point-like nucleus approximation overestimates the muonic and hadronic vacuum polarization corrections to the $g$-factors by more than two orders of magnitude for heavy elements. In the point-like nucleus approximation, the hadronic vacuum polarization corrections also appear to be larger than the muonic vacuum polarization corrections for heavy elements (compare Tables \ref{muonicVP} and \ref{hadrVPTable}), which is due to the hadronic Uehling potantial for point-like nucleus being larger in magnitude than the muonic Uehling potential in the small $r$ region (see e.g. Figure 1 in Ref \cite{Breidenbach2022}), but our results confirm that for extended nuclei, muonic vacuum polarization corrections are strictly larger than the hadronic vacuum polarization corrections for all elements (see Figures \ref{delta_E_figure} and \ref{delta_g_figure}).

Following the results we obtained, it becomes clear that the dependece on nuclear charge distribution model significantly affects the properties of bound muons, and it is necessary to construct precise theoretical models for the nucleus in order to be able to have accurate theoretical predictions for the properties of the bound muon. A significant contributor that was not considered in this work is the nuclear polarization effect, which has been studied for electronic atoms \cite{Debierre2022,Patkos2023} and for the energy levels of bound muons \cite{Kalinowski2018,Valuev2022}, but remains undetermined for the $g$-factors.

Among the QED effects, the self-energy corrections and the magnetic-loop vacuum polarization corrections to the $g$-factors were not considered in this work. Both of these effects have been studied heavily in the point-nucleus case (such as in \cite{Beier2000,Lee2005,Yerokhin2017}), but it becomes significantly more difficult to account for the finite size of the nucleus in those effects. Nuclear model dependence of the self-energy correction to the energy levels of muonic atoms has recently been studied in \cite{Oreshkina2022}, but the finite-nuclear-size self-energy corrections to the $g$-factors have still not been worked out.

\begin{acknowledgments}
A.Ç. would like to thank the Theoretical Quantum Dynamics and Quantum Electrodynamics division of the Max Planck Institute for Nuclear Physics for their hospitality during his stay and internship in Heidelberg, where this work was completed. Supported by the Deutsche Forschungsge-
meinschaft (DFG, German Research Foundation) Project-ID
273811115 - SFB 1225.
\end{acknowledgments}

\bibliography{refs}

\end{document}